\documentclass[aps,prb,twocolumn,10pt,amsmath,amssymb,floatfix,superscriptaddress]{revtex4-2}

\usepackage{graphicx} 
\usepackage{bm} 
\usepackage{hyperref} 
\usepackage{braket,bbold,color}
\usepackage{hyperref}

\begin{document}

\title{Interaction-induced crystalline topology of excitons}

\author{Henry Davenport}
\affiliation{Blackett Laboratory, Imperial College London, London SW7 2AZ, United Kingdom}

\author{Johannes Knolle}
\affiliation{Technical University of Munich, TUM School of Natural Sciences, Physics Department, 85748 Garching, Germany}
\affiliation{Munich Center for Quantum Science and Technology (MCQST), Schellingstr. 4, 80799 M\"unchen, Germany}
\affiliation{Blackett Laboratory, Imperial College London, London SW7 2AZ, United Kingdom}

\author{Frank Schindler}
\affiliation{Blackett Laboratory, Imperial College London, London SW7 2AZ, United Kingdom}

\begin{abstract}
We apply the topological theory of symmetry indicators to interaction-induced exciton band structures in centrosymmetric semiconductors. Crucially, we distinguish between the topological invariants inherited from the underlying electron and hole bands, and those that are intrinsic to the exciton wavefunction itself. Focusing on the latter, we show that there exists a class of exciton bands for which the maximally-localised exciton Wannier states are shifted with respect to the electronic Wannier states by a quantised amount; we call these excitons \emph{shift excitons}. Our analysis explains how the exciton spectrum can be topologically nontrivial and sustain exciton edge states in open boundary conditions even when the underlying noninteracting bands have a trivial atomic limit. We demonstrate the presence of shift excitons as the lowest energy neutral excitations of the Su-Schrieffer-Heeger model in its trivial phase when supplemented by local two-body interactions, and show that they can be accessed experimentally in local optical conductivity measurements.
\end{abstract}

\maketitle

\section{Introduction}

The notion of topological insulators has revolutionised our understanding of electronic properties in condensed matter systems~\cite{HasanKaneRMP10, QSHEffectGrapheneMele2005, QSHEffectBernevig2006}. Among these, topological crystalline insulators (TCIs)~\cite{FuTCI11, topocrysInsulatorsRobertJan2013} stand out due to their reliance on crystalline symmetries, such as mirror, rotational, and inversion symmetries, to protect gapless surface and hinge states. The classification of TCIs has significantly broadened the landscape of topological matter and can by now be considered quite mature~\cite{RobertJanPRX17, Ashvin230SINatComm17, AndreiTQC17, HermeleTopoCrystalSciAdv19, ShiozakiSpectralSequencePRB22}. This is especially true for the important subclass of \emph{symmetry-indicated} topological bands that can be distinguished by their space group symmetry representations at high-symmetry momenta in the Brillouin zone.

As the frontier of topological materials expands, there is a growing interest in generalising the framework of band topology to incorporate electronic interactions. Most research efforts have been directed towards understanding the effects of interactions on ground states, with a particular recent focus on classifying TCIs in an interacting context~\cite{MartinaInteractingTQC23,ShiozakiSPTClassification23,NarenManjunathTCIClassification24,JonahInteractingTQCNatComm24}. Beyond ground states, the study of interaction-induced \emph{excitations}, particularly excitons, represents an exciting and less explored avenue. 

Excitons, bound states of electrons and holes, arise from the electron-hole continuum as a result of interactions. Unlike the band insulating ground state, the exciton band structure can be modified substantially by interactions even with characteristic energy much less than the insulating gap. This characteristic makes them a promising candidate for exploring new topological phenomena induced by interactions in real materials. Towards this goal, we here apply the topological theory of symmetry indicators to exciton band structures in centrosymmetric semiconductors.

A key aspect of our study is distinguishing between topological invariants inherited from the underlying electron and hole bands and those intrinsic to the exciton wavefunction itself. While much previous work has concentrated on the former -- for instance, by investigating excitons in Chern bands \cite{ChenPhysRevB2017, FuChernExcitons, xie2024theory, ParamesaranPhysRevLet2021, QSHInsulatorExcitonsFabrizio} -- we here focus on the latter and introduce the concept of \emph{shift excitons}. These bulk excitons exhibit maximally-localised \emph{exciton Wannier states}~\cite{NeatonExcitonMLWFsPRB23} that are shifted relative to the electronic Wannier states by a quantised amount. This leads to nontrivial exciton band structures that can support exciton edge states even when the noninteracting electronic bands are completely trivial, offering a new avenue to stabilise robust topological features in condensed matter.

We show that shift excitons can be the lowest energy neutral excitations in the Su-Schrieffer-Heeger (SSH) model~\cite{SSHPRL79} \emph{in its trivial phase}, when supplemented by suitable strictly local two-body interactions. Furthermore, we derive how shift excitons can be experimentally accessed through local optical conductivity measurements~\cite{localOpticCondExp, topExcitonsMoire}. 

We now briefly outline the paper. In Sec.~\ref{sec: TheorySection} we introduce the general concept of shift excitons. Then, in Sec.~\ref{SSHSection} we show how shift excitons emerge in an interacting generalisation of the SSH model. Finally, in Sec.~\ref{sec: experimental consequences}, we discuss local optical conductivity measurements to diagnose shift excitons. We summarise our findings and lay out future directions in Sec.~\ref{sec: discussion}.

\section{Theory of shift excitons}
\label{sec: TheorySection}

We here introduce shift excitons for centrosymmetric semiconductors in one spatial dimension (1D). For clarity we focus on a spinless two-band tight-binding model at half filling, \emph{i.e.}, the ground state has only the lower-energy band occupied; the $n$ band generalisation in $d$ dimensions is straightforward. We assume that the occupied band is fully gapped from the empty band at all momenta and that both bands realise an \emph{unobstructed atomic limit}~\cite{AndreiTQC17}, \emph{i.e.}, their Wannier centers are located at the atomic ions at the origin of the unit cell and have zero polarisation (we relax this assumption in App.~\ref{apdx:ShiftDefinition}). The electronic bands then have the same inversion $\mathcal{I}$ eigenvalues at both of the high symmetry points $k = 0,\pi$ in the 1D Brillouin zone (BZ) as long as the inversion center is chosen to be at the center of the unit cell~\cite{AlexandradinataWilsonPRB14}. We now study the resulting exciton bands in the presence of interactions.

Let $c_{k, \mathrm{occ}}$ ($c_{k, \mathrm{emp}}$) annihilate an electron in the occupied (empty) band at crystal momentum $k \in [0,2\pi)$. The noninteracting band insulator ground state is given by $\ket{GS} = \prod_{k} c^\dagger_{k, \mathrm{occ}} \ket{0}$, where $\ket{0}$ is the fermionic vaccuum. The low energy exciton spectrum can be found by projecting the system's Hamiltonian into the variational exciton basis $c^\dagger_{p+k, \mathrm{emp}} c_{k, \mathrm{occ}} \ket{GS}$, where $p$ is the total momentum of the electron-hole pair that is conserved due to translational symmetry, and $k$ cycles through different relative momenta. 

The exciton eigenstate for a given exciton band at total momentum $p$ has the form 
\begin{equation} \label{eq: exciton basis state}
\ket{\phi^{p}} = \sum_k \phi^{p}_k c^\dagger_{p+k, \mathrm{emp}} c_{k, \mathrm{occ}} \ket{GS},
\end{equation}
where we call $\phi^{p}_k \in \mathbb{C}$ the exciton wavefunction. Crucially, this expansion of $\ket{\phi^{p}}$ makes it clear that the exciton state may have topological invariants contributed by the noninteracting Bloch states, which enters the exciton state in Eq.~\eqref{eq: exciton basis state} via the electron and hole operators $c^\dagger_{p+k, \mathrm{emp}}$, $c_{k, \mathrm{occ}}$, and/or the exciton wavefunction itself; we here focus on the latter case to establish a rigorous bulk-boundary correspondence. For instance, if the electrons of a two-dimensional system have a nonzero Chern number, a Coulomb interaction may give rise to chiral exciton edge states that are directly inherited from the electronic edge states~\cite{ChenPhysRevB2017, FuChernExcitons, xie2024theory, ParamesaranPhysRevLet2021, QSHInsulatorExcitonsFabrizio}, nevertheless, the bulk excitons of the same system may well remain trivial or even gapless.

As a basic and simple example of intrinsic exciton topology, we here investigate the crystalline topology of exciton bands using the symmetry indicators of $\mathcal{I}$ symmetry~\cite{AlexandradinataWilsonPRB14,RobertJanPRX17,Ashvin230SINatComm17,AndreiTQC17}. This allows us to differentiate between topologically distinct types of exciton bands using the $\mathcal{I}$ eigenvalues of the exciton states at the high symmetry points $\tilde{p} = 0,\pi$ of the BZ. 

We first choose a convenient gauge for the Bloch states $\ket{\psi^{p, \alpha}}$ associated with the underlying noninteracting bands, such that $c^\dagger_{p, \alpha}= \sum_{i} \braket{p, i | \psi^{p, \alpha}} c^\dagger_{p, i}$, $\alpha \in \{\mathrm{occ}, \mathrm{emp}\}$, and $c^\dagger_{p, i}$ creates a Bloch wave with wavevector $p$ on sublattice/orbital $i$ in the unit cell. 
We adopt the gauge $\ket{\psi^{-p, \alpha}} = \lambda^{\alpha}_I U_I \ket{\psi^{p, \alpha}}$. Here, the unitary matrix $U_I$ represents $\mathcal{I}$ symmetry in the single-particle Hilbert space, and $\lambda^{\alpha}_I$ denotes the $\mathcal{I}$ eigenvalue of the noninteracting band at both high-symmetry points $\tilde{p} = 0,\pi$, which are equal by our assumption of trivial (unobstructed) electronic bands. (Recall that since $\mathcal{I}^2 = 1$, any $\mathcal{I}$ eigenvalue can only take values $\pm 1$.) 

It follows that $\hat I c^\dagger_{p, \alpha} \hat I^\dagger = \lambda^{\alpha}_I c^\dagger_{-p, \alpha}$, where $\hat I$ represents $\mathcal{I}$ symmetry in the many-body Hilbert space. Consequently, at $\tilde p = 0,\pi$, the $\mathcal{I}$ eigenvalue can be separated into four distinct contributions:
\begin{align}
 \hat I \ket{\phi^{\tilde p}} &= \big(\lambda_I^{\mathrm{occ}\vphantom{\mathrm{emp}}}\lambda_I^{\mathrm{emp}}
\lambda_I^{GS}\big)\sum_k \phi^{\tilde p}_{-k} c^\dagger_{\tilde p+k, \mathrm{emp}} c_{k, \mathrm{occ}} \ket{GS} \nonumber \\
&\equiv\big(\lambda_I^{\mathrm{occ}\vphantom{\mathrm{emp}}}\lambda_I^{\mathrm{emp}}
\lambda_I^{GS}\big) \lambda^{\mathrm{exc}}_I(\tilde p) \ket{\phi^{\tilde p}}.
\end{align}
There are contributions from the $\mathcal{I}$ eigenvalue of the ground state ($\lambda^{GS}_I$) as well as the underlying electronic bands. However, the only contribution which varies between different high symmetry points is the excitonic contribution $\lambda^{\mathrm{exc}}_I(\tilde p)$ that comes from $\phi^{\tilde p}_{k} = \lambda^{\mathrm{exc}}_I(\tilde p) \phi^{\tilde p}_{-k}$. We refer to an exciton wavefunction with $\lambda^{\mathrm{exc}}_I(0) = \lambda^{\mathrm{exc}}_I(\pi)$ as trivial, otherwise we call the exciton wavefunction nontrivial. Nontrivial excitons can therefore arise even in a trivial single-particle band structure, as long as the $\mathcal{I}$ eigenvalues of the exciton wavefunction differ at $\tilde p$. Just like for electronic Wannier states, the Wannier centres of the maximally localised \emph{exciton Wannier states}~\cite{NeatonExcitonMLWFsPRB23} then shift by a quantised amount~\cite{AlexandradinataWilsonPRB14}.

To demonstrate the exciton shift, we define the excitonic Wannier state $\ket{W^{R'}}$ centered at position $R'$, $\ket{W^{R'}} = \frac{1}{\sqrt{L}}\sum_{p} e^{-ipR'} \ket{\phi^{p}}$. We assume that a smooth gauge (as a function of $p$) has been found for the exciton wavefunction $\phi^{p}_k$ so that the $\ket{W^{R'}}$ are exponentially localised in space; in 1D this is always possible~\cite{MarzariVanderbiltSmoothGaugePRB97}. The resulting state can be expanded as
\begin{equation} \label{eq: exciton Wannier state}
\ket{W^{R'}} = \sum_{R, \Delta} W^{R-R'}_{\Delta} c^\dagger_{R, \mathrm{emp}} c_{R-\Delta, \mathrm{occ}}\ket{GS}.
\end{equation}
Here, $c^\dagger_{R, \alpha} = \frac{1}{\sqrt{L}} \sum_p e^{-i p R} c^\dagger_{p, \alpha}$ are the maximally localised electronic Wannier states associated with the band $\alpha \in \{\mathrm{emp},\mathrm{occ}\}$ in the unit cell at $R$, and 
\begin{equation}
W^{R}_{\Delta} = \frac{1}{L\sqrt{L}} \sum_{p, k} e^{ipR} e^{ik\Delta} \phi^{p}_k.
\label{eq:ExcitonWannierFunction}
\end{equation}

We define the exciton shift for the exciton Wannier state at unit cell $R'=0$ as 
\begin{equation}
s_{\mathrm{exc}} = \sum_{R, \Delta} |W^{R}_\Delta|^2 R. 
\end{equation}
While we have so far assumed that the electronic Wannier centers are $x_\mathrm{occ} = x_\mathrm{emp} = 0$, we show in App.~\ref{apdx:ShiftDefinition} that in general
\begin{equation}
s_{\mathrm{exc}} = \braket{W^{0}| \hat{x}^{(\mathrm{e})}_\mathrm{emp} | W^{0}} - x_\mathrm{emp}
= \braket{W^{0}| \hat{x}^{(\mathrm{h})}_\mathrm{occ} | W^{0}} - x_\mathrm{occ},
\end{equation}
where $\hat{x}^{(\mathrm{e})}_\mathrm{emp} = \sum_R (R+x_\mathrm{emp}) c^\dagger_{R, \mathrm{emp}} c_{R, \mathrm{emp}}$ [$\hat{x}^{(\mathrm{h})}_\mathrm{occ} = \sum_R (R+x_\mathrm{occ}) c_{R, \mathrm{occ}} c^\dagger_{R, \mathrm{occ}}$] is the empty-band electron (occupied-band hole) projected position operator~\cite{MarzariVanderbiltSmoothGaugePRB97}. Correspondingly, the electron and hole making up the exciton Wannier state are shifted on average by the same amount $s_{\mathrm{exc}}$ with respect to the noninteracting electron and hole Wannier centers, respectively. Equivalently, $s_{\mathrm{exc}}$ represents the shift with which the actual exciton center of mass is offset from the ``naive" exciton center of mass that one would obtain by simply pairing up electron and hole Wannier states in each unit cell. We note that the relative nature of the exciton shift $s_{\mathrm{exc}}$ implies that it cannot be trivialised by a change of unit cell.
\begin{figure}[t]
\centering
\includegraphics[width=0.48\textwidth]{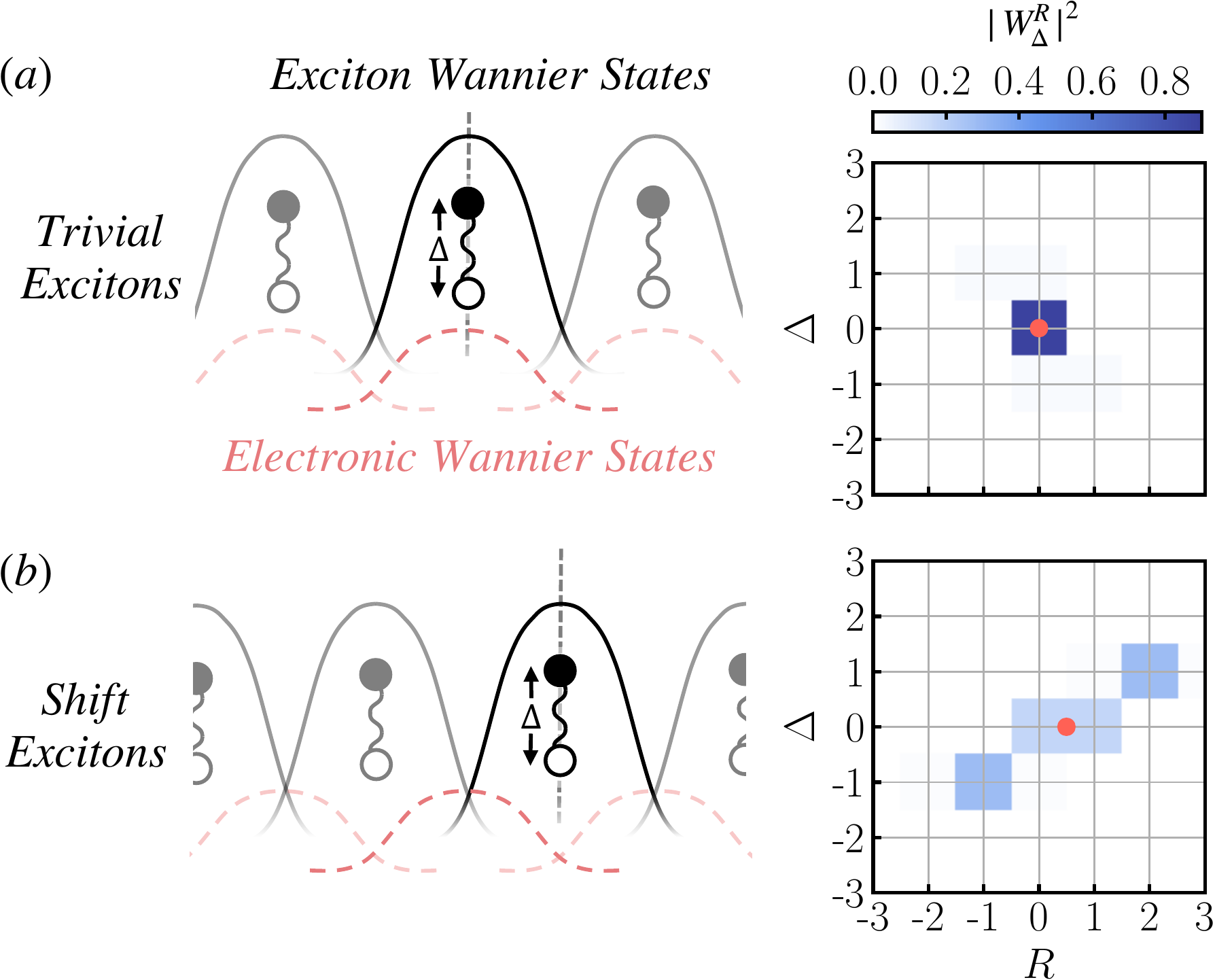}
\caption{Conceptual diagram of the exciton Wannier states for (a) trivial and (b) shift excitons. The right panels show examples of explicitly calculated exciton Wannier states in the SSH model described in Sec.~\ref{SSHSection} as a function of absolute position $R$ and relative spread $\Delta$. The inversion center for the Wannier states is shown as a red dot, indicating a nontrivial exciton shift $s_{\mathrm{exc}} = 1/2$ for panel b.}
\label{fig: theory}
\end{figure}
We next demonstrate that the excitonic shift is quantised to $s_{\mathrm{exc}} = 0,\frac{1}{2}$ in centrosymmetric systems. $\mathcal{I}$ symmetry requires that $\phi^{p}_k = e^{i \alpha(p)}\phi^{-p}_{-k}$, where $e^{i \alpha(p)}$ is a phase factor (the ``sewing matrix"~\cite{AlexandradinataWilsonPRB14}) that can always be chosen to be a smooth function of $p$ in 1D. The $\mathcal{I}$ eigenvalues at the high symmetry points constrain the form of $\alpha(p)$. For trivial exciton bands we have $\lambda^{\mathrm{exc}}_{I}(\tilde p) = \lambda^{\mathrm{exc}}_{I}$, and hence the gauge $\alpha(p)$ can be chosen constant i.e. $\alpha(p) = \alpha$ where $e^{i\alpha} = \lambda^{\mathrm{exc}}_I$. From Eq.~\eqref{eq:ExcitonWannierFunction}, this leads to $W^{R}_{\Delta} = \lambda^{\mathrm{exc}}_I W^{-R}_{-\Delta}$ and hence the shift becomes
\begin{equation}
\begin{aligned}
s_{\mathrm{exc}} &=\sum_{R, \Delta} |W^{R}_\Delta|^2 R
=\sum_{R, \Delta} |\lambda^{\mathrm{exc}}_I W^{-R}_{-\Delta}|^2 R\\
&=-\sum_{R, \Delta} |W^{R}_{\Delta}|^2 R=0.
\end{aligned}
\end{equation}
However for nontrivial exciton bands, the $\mathcal{I}$ eigenvalues differ at the high symmetry points. The gauge therefore must be $\alpha(p) = \alpha -p$ where $\alpha$ is defined via $e^{i\alpha} = \lambda^{\mathrm{exc}}_I(\tilde p = 0)$. This gauge choice leads to $W^{R}_{\Delta} = \lambda^{\mathrm{exc}}_I(\tilde p = 0) W^{-(R-1)}_{-\Delta}$~and hence
\begin{equation}
\begin{aligned}
s_{\mathrm{exc}} &=\sum_{R, \Delta} |\lambda^{\mathrm{exc}}_I(\tilde p = 0) W^{-(R-1)}_{-\Delta}|^2 R\\
&=-\sum_{R, \Delta} |W^{R}_{\Delta}|^2 (R-1) 
= -s_{\mathrm{exc}} + 1 =\frac{1}{2}.
\end{aligned}
\label{eq:WanniershiftExcitons}
\end{equation}
Fig.~\ref{fig: theory} demonstrates the difference between shift excitons and trivial excitons and shows some explicitly calculated exciton Wannier states for the model introduced in Sec.~\ref{SSHSection}.

Importantly, a nonzero exciton shift $s_{\mathrm{exc}} = 1/2$ implies the presence of a counting mismatch of exciton eigenstates between periodic boundary conditions (PBC) and open boundary conditions (OBC), in analogy to the filling anomaly of obstructed atomic limits~\cite{BenalcazarCnCornerStatesPRB19}. For this, it is crucial to consider an OBC termination that \emph{does not cut through any single-particle Wannier states}. (Compare this with the usual condition in TCIs that the OBC termination not cut through unit cells.) Given a lattice of $N$ unit cells, there will then be $N$ exciton states that form a band in PBC. If and only if this band has a nontrivial shift $s_{\mathrm{exc}} = 1/2$, it will consist of $N \pm 1$ contiguous exciton states in OBC; potentially (in the case of $N-1$ states) leaving behind $2$ mid-gap exciton states localised at opposite edges of the system. Since these edge states are exponentially localised and exchanged by $\mathcal{I}$ symmetry, they must remain degenerate. In particular, merging them with the bulk excitons does not remove the counting mismatch $N \pm 1 \neq N$ with respect to PBC. 

Next, we will show that it is possible to stabilise shift excitons and their associated edge states in a minimal tight-binding model with local interactions.

\section{Shift excitons in the SSH model}
\label{SSHSection}

\begin{figure}[t]
\centering
\includegraphics[width=0.48\textwidth]{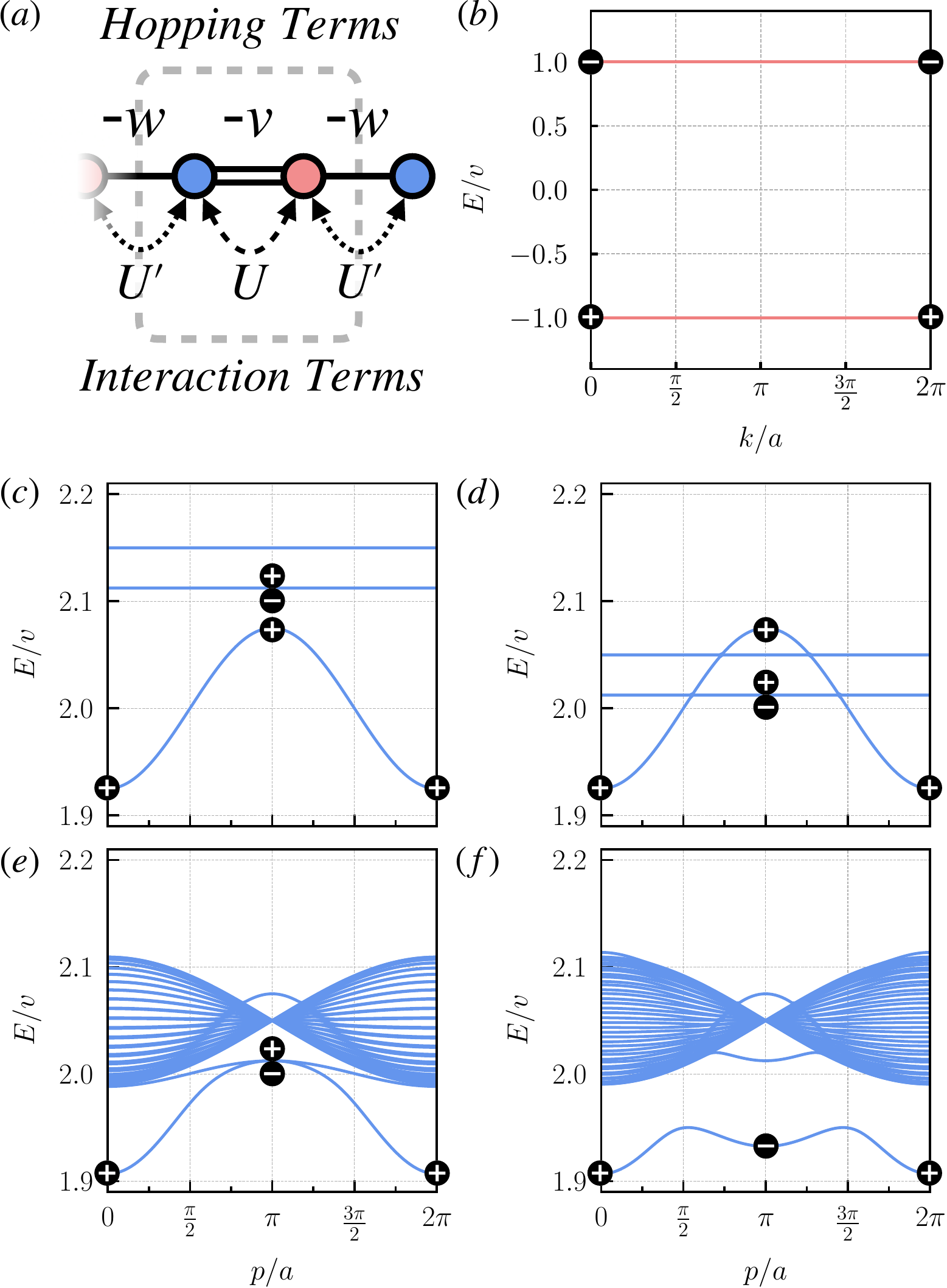}
\caption{$(a)$ The interacting SSH model with hoppings $-v, -w$ between the two sublattices $A$ (blue) and $B$ (red) and interactions $U, U'$ between electrons on the two sublattices. $(b)$~The noninteracting band structure of the SSH model in the trivial phase in the flat-band limit  ($v=1, w=0$). The $\mathcal{I}$ eigenvalues are marked at the high symmetry momenta. $(c)-(f)$ show the construction of a nontrivial exciton band in the interacting SSH model by hybridising trivial bands. The relevant $\mathcal{I}$ eigenvalues are marked at the high symmetry momenta. In $(c)$ the lowest energy flat band is twice degenerate so is marked with two $\mathcal{I}$ eigenvalues. All spectra have $v=1$ the remaining parameters are (c) $U=0.15$, $U'=0.15$, $w=0$, (d) $U=0.05$, $U'=0.15$, $w=0$, (e) $U=0.05$, $U'=0.15$, $w=0.03$, (f) $U=0.05$, $U'=0.15$, $w=0.03$ and pair hopping $V=0.01$.}
\label{fig: ssh model 1}
\end{figure}

Shift excitons can be realised in an interacting generalisation of the spinless 1D centrosymmetric SSH model (see Fig.~\ref{fig: ssh model 1}a) \emph{in its trivial phase} -- that is, there are no single particle edge states in open boundary conditions (OBC) when terminating with full unit cells. In our conventions, this model has two sublattices labelled $A$ and $B$ and a hopping $(-v)$ between sites within the unit cell and $(-w)$ between adjacent unit cells. $\mathcal{I}$ symmetry inverts about the center of the unit cell, so that it exchanges $A$ and $B$. The Coulomb interaction introduces quartic terms into the tight-binding Hamiltonian. The specific form of these quartic terms is determined by the underlying atomic orbitals, but initially, we focus solely on the contributions from the intra- and inter-unit cell extended Hubbard interactions denoted $U$, $U'$ (these are the simplest possible interaction terms due to the spinless nature of the model):
\begin{equation} \label{eq: SSH Hamiltonian}
\begin{aligned}
\hat H =& - v\sum_R (c^\dagger_{R, A} c_{R, B} + c^\dagger_{R, B} c_{R, A})\\& - w\sum_R (c^\dagger_{R, B} c_{R+1, A} + c^\dagger_{R+1, A} c_{R, B})\\& + U\sum_R n_{R, A} n_{R, B} + U' \sum_R n_{R, B} n_{R+1, A}.
\end{aligned}
\end{equation}
We first consider the system in periodic boundary conditions (PBC) with $v$ set to unity and $w,U,U'$ treated perturbatively on top of the noninteracting ground state of the dimerised SSH model (i.e., the ground state at $v=1, w=U=U'=0$). Since the inter-unit cell hopping $w$ is set to zero, there are clearly no edge states in OBC when terminating with full unit cells. Correspondingly, the noninteracting part of the model is in the trivial phase both with respect to chiral symmetry~\cite{AsbothBook16} and $\mathcal{I}$ symmetry: since we can localise the sublattices $A$ and $B$ at the center of the unit cell without loss of generality, it realises an \emph{unobstructed} atomic insulator~\cite{AndreiTQC17}. This corresponds to an unperturbed ground state $\ket{GS} = \prod_R c^\dagger_{R, +} \ket{0}$ where $c^\dagger_{R, +}$ is the creation operator for the Wannier state at unit cell $R$ for the filled electronic band. In the dimerised limit, the Wannier states of the occupied ($+$) and empty ($-$) bands are compact and strictly localised to a single unit cell~\cite{NoncompactPRB21}:
\begin{equation}
c^\dagger_{R, \pm} = \frac{c^\dagger_{R, A} \pm c^\dagger_{R, B}}{\sqrt{2}}, \quad \hat{I} c^\dagger_{R, \pm} \hat{I}^\dagger = \pm c^\dagger_{R, \pm}.
\label{eq:electronWannierStatesSSH}
\end{equation}
Excitons in this model consist of excitations of electrons from the occupied Wannier states ($\mathcal{I}$ eigenvalue $+1$) to the empty Wannier states ($\mathcal{I}$ eigenvalue $-1$). We therefore project the Hamiltonian in Eq.~\eqref{eq: SSH Hamiltonian} into the variational exciton basis $c^\dagger_{R+\Delta, -} c_{R, +} \ket{GS}$ to give an effective exciton Hamiltonian matrix $H_{R, R', \Delta, \Delta'}$. This matrix describes the scattering between excitons centered at $R$, with electron-hole separation $\Delta$, to those located at $R'$, with electron-hole separation $\Delta '$. The Hamiltonian is translationally invariant, $H_{R, R', \Delta, \Delta'} = H_{R-R', 0, \Delta, \Delta'}$, hence we can perform a Fourier transform resulting in an effective exciton Hamiltonian $H_{\Delta, \Delta'}(p)$ that depends on total momentum $p$ and takes the form
\begin{equation}
\begin{aligned}
H_{\Delta, \Delta'}(p) = \delta_{\Delta, \Delta'} \bigg[ 2v + U\delta_{\Delta \neq 0} + \frac{U'}{4} (\delta_{\Delta \neq 1}+\delta_{\Delta \neq -1}-2)\\  - \frac{U'}{2} \delta_{\Delta,0} \cos p \bigg] - 
\frac{w}{2}(\delta_{\Delta', \Delta+1}+\delta_{\Delta', \Delta-1})\left[1+e^{i p (\Delta' - \Delta)}\right],
\label{eq:EffExcitonHam}
\end{aligned}
\end{equation}
where we have abbreviated $\delta_{\Delta \neq x} \equiv 1-\delta_{\Delta,x}$.
The exciton spectrum above the ground state for the choice $U=U'$ and $w=0$ is shown in Fig.~\ref{fig: ssh model 1}c and exhibits a trivial dispersive band with lowest energy $E = v -\frac{U'}{4}$ that is gapped from two sets of degenerate flat bands at higher energy. The lower set of flat bands is doubly degenerate and corresponds to bound states with electron-hole separation $\Delta=1$ and $\Delta = -1$. For a system with $L$ unit cells, the higher set of flat bands consists of $L(L-3)$ states forming the electron-hole continuum, which is flat in the dimerised limit.

To construct a nontrivial exciton band, we begin by considering the effect of further lowering the energy of the doubly degenerate flat band by decreasing the interaction strength $U$ with respect to $U'$ as shown in Fig.~\ref{fig: ssh model 1}d. These flat bands should hybridise with the dispersive band when the inter-unit cell hopping term $w$ is turned on. While this may result in a nontrivial exciton band (with opposite $\mathcal{I}$ eigenvalues at $p=0,\pi$), as shown in Fig.~\ref{fig: ssh model 1}e, this band is not yet gapped at $p=\pi$. To understand why this is in fact not possible, we consider a projected exciton Hamiltonian $H'(p)$ that is restricted to $\Delta = 1, 0, -1$. The spectrum of this restricted Hamiltonian gives a good approximation for the low energy physics, because the 3 lowest energy bands in the exciton spectrum are all bound states and so are exponentially localised in $\Delta$. In the basis $\Delta \in \{1, 0, -1\}$, we obtain $H'(p)=$
\begin{equation}
    \begin{pmatrix}
    2v +U - \frac{U'}{4} & -\frac{w}{2}(1+e^{-ip}) & 0 \\
    -\frac{w}{2}(1+e^{ip}) & 2v-\frac{U'}{2} \cos(p) & -\frac{w}{2}(1+e^{-ip})\\
    0 & -\frac{w}{2}(1+e^{ip}) & 2v +U - \frac{U'}{4}
    \end{pmatrix}.
    \label{eq:3x3Matrix}
\end{equation}
Clearly, at momentum $p=\pi$, the term multiplying $w$ vanishes, implying that the negative $\mathcal{I}$ eigenstate $(1, 0, -1)^T$ and the positive $\mathcal{I}$ eigenstate $(1, 0, 1)^T$ must have the same energy.

Although the gap at $p=\pi$ can be opened using longer-range hopping terms, we aim to not only obtain a nontrivial band gapped at any given momentum, but also to fully gap the nontrivial band across all momenta such that the gap persists in OBC. We show below that when this occurs edge localised excitons can exist in the band gap. In order for the gap to remain in OBC, however, the highest energy state in the nontrivial band must be lower in energy than all states in the remaining bands. This requirement can never be achieved in the spectrum of the reduced Hamiltonian in Eq.~\eqref{eq:3x3Matrix}, and we show in App.~\ref{apdx:FullyGapped} that adding longer range hopping or further Hubbard-type interaction terms (as described in App.~\ref{apdx:GeneralTheory}) also cannot fully gap the nontrivial band. 

Generically, however, the Coulomb interaction projected into tight binding models leads to quartic terms beyond terms of the Hubbard type. For example one can also obtain pair hopping terms and these terms can be used to fully open the exciton gap. Consider for instance the pair hopping term
\begin{equation} \label{eq: full gapping term}
\begin{aligned}
\hat H_V = V \sum_{ R} (c^\dagger_{ R + 1, B} c_{ R, A} c^\dagger_{ R-1, A} c_{ R, B}\\ - c^\dagger_{ R-1, B} c_{ R, A} c^\dagger_{ R-1, A} c_{ R, B})+\mathrm{h.c.},
\end{aligned}
\end{equation}
which when projected into the effective $3\times 3$ exciton Hamiltonian adds the term
\begin{equation}
H'_V(p)=
\begin{pmatrix}
4V \cos(p) & 0 & -4V e^{-ip} \\
0 & 0 & 0\\
-4V e^{ip} & 0 & 4V \cos(p)
\end{pmatrix}.
\end{equation}
As a consequence of adding this term with $V>0$, the energy of odd-$\mathcal{I}$ states at $p=0$ is increased whilst it is decreased at $p=\pi$. For reasonable parameter choices, this term opens up the desired gap at all momenta as shown in Fig.~\ref{fig: ssh model 1}f. In App.~\ref{apdx:FlatBand} we show that in addition to being able to gap out a nontrivial exciton band, it is possible to make it completely flat, potentially leading to strongly correlated exciton condensate states.

\begin{figure}[t]
\centering
\includegraphics[width=0.5\textwidth]{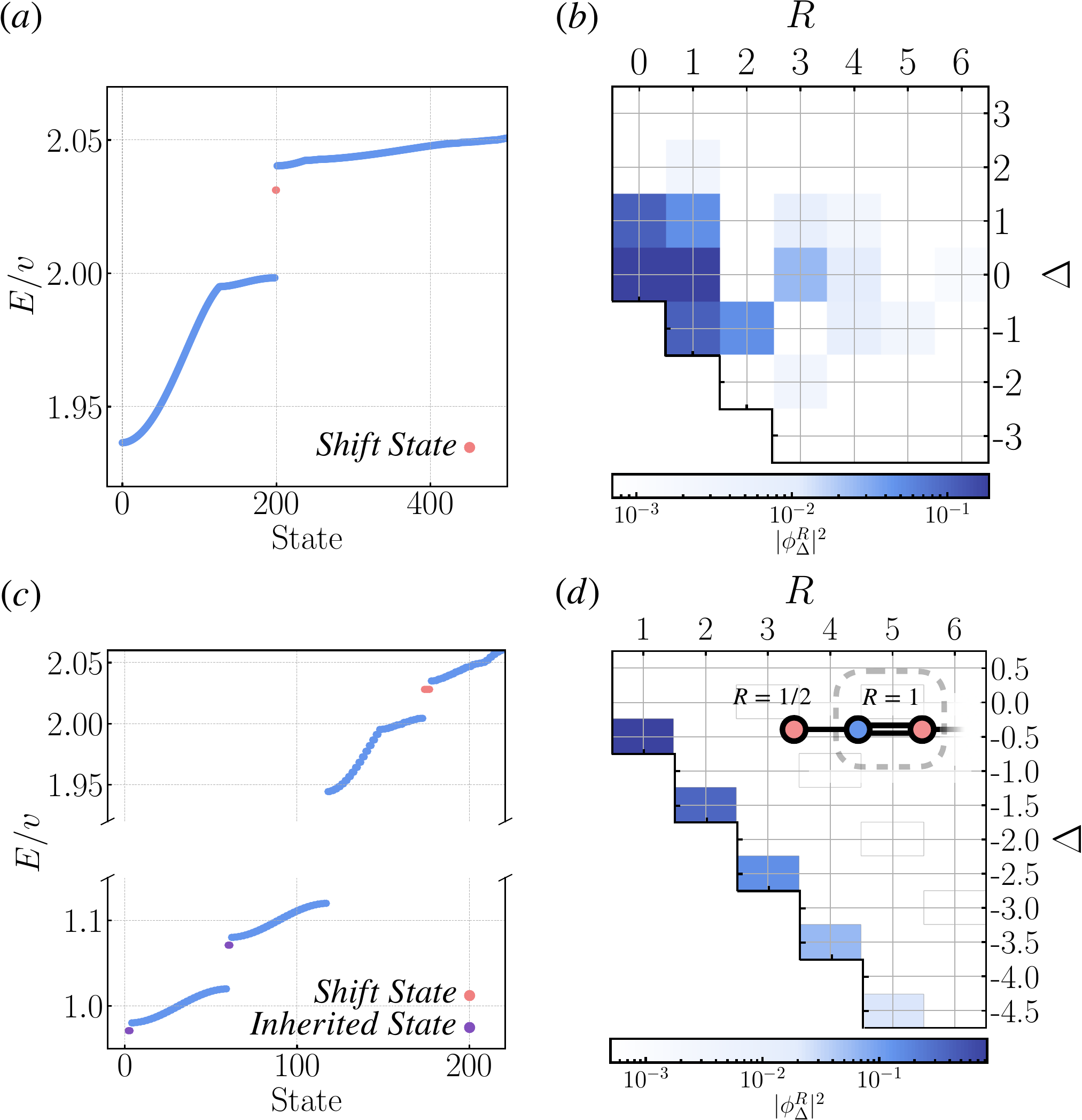}
\caption{$(a)$ Spectrum in OBC with the chain terminated on strong bonds (\emph{i.e.}, on a full unit cell when $|v| > |w|$) with $v=1.0$, $w=0.03$, $U=0.1$, $U'=0.1$ and $V=0.01$. The exciton states in our variational basis are $\sum_{R, \Delta} \phi^R_\Delta c^\dagger_{R+\Delta, -} c_{R,+} \ket{GS}$. The mod squared of the wavefunction ($|\phi^R_\Delta|^2$) for one of the interaction-induced shift exciton edge states (marked Shift State in $(a)$) is shown in $(b)$. The cut-out removes disallowed states (\emph{e.g.} $\Delta < -R$ is impossible because the left edge of the system is at $R=0$). $(c)$ Spectrum in OBC with the chain instead terminated on a weak bond (\emph{i.e.}, cutting through the unit cell when $|v| > |w|$). All parameters but $w$, now $w=0.02$, are the same as in $(a)$ above. For these parameters, there are exciton edge states which are inherited from the noninteracting edge states, \emph{e.g.} see $(d)$.  The single site added to the left end is labelled $R=\frac{1}{2}$.  Excitons which excite electrons from the bulk to the $R=\frac{1}{2}$ site (if it is unoccupied) therefore involve half integer~$\Delta$.}
\label{fig: ssh model 2}
\end{figure}

The nontrivial shift $s_{\mathrm{exc}} = 1/2$ of the exciton Wannier centers that follows from their opposite inversion eigenvalues at $\tilde{p} = 0, \pi$ gives rise to a bulk-boundary correspondence as explained at the end of Sec.~\ref{sec: TheorySection}. We find that when the $|v| > |w|$ chain is terminated with full unit cells (and not cutting through any single-particle Wannier states), there are no edge states in the noninteracting SSH model and yet the interaction-induced nontrivial topology of the exciton bands leads to exciton edge states. The spectrum and profile of a mid-gap exciton edge state at the left edge of the chain is shown in Fig.~\ref{fig: ssh model 2}a,b.

It is educating to note that the interaction-induced exciton edge states discussed here -- a direct consequence of bulk shift excitons -- are fundamentally different in character from the exciton states that can arise from \emph{noninteracting} bulk topology. In 1D TCIs, single-particle edge states appear -- at least in the guise of a filling anomaly~\cite{BenalcazarCnCornerStatesPRB19} -- generically when the system is terminated on a Wannier center, and this may lead to exciton edge states \emph{inherited} from the single-particle edge states. However, as explained at the end of Sec.~\ref{sec: TheorySection}, any such termination that cuts through single-particle Wannier states does not give rise to a proper bulk boundary correspondence for excitons -- the OBC exciton spectrum will not have a well-defined state counting mismatch with PBC and instead depends sensitively on the way the system is terminated. Nevertheless, we here study also the inherited edge excitons arising from a nontrivial termination of the SSH model, as that is the usual setting in which the SSH model is studied and where it hosts its celebrated single-particle edge states at zero energy. Specifically, the chain is terminated on a weak bond, this is equivalent to choosing $w > v$ and then again terminating with full unit cells. As a consequence of the Hubbard-type interaction terms, and irrespective of whether the bulk excitons have shift $s_{\mathrm{exc}} = 0$ or $s_{\mathrm{exc}} = 1/2$, the extra site at each edge of the chain then hosts inherited exciton bound states that decay exponentially into the bulk (e.g. see Fig.~\ref{fig: ssh model 2}c,d). Not only do these inherited edge states occur at a very different energy $E \sim v$ (compare with the shift exciton edge states that appear at $E \sim 2v$), but they also differ significantly in their structure: they have a characteristic diagonal shape in the $(R,\Delta)$ diagram shown in Fig.~\ref{fig: ssh model 2}d, not present for the interaction-induced edge states (Fig.~\ref{fig: ssh model 2}b). The non-zero elements of the inherited edge exciton wave function [of the form Eq.~\eqref{eq: exciton basis state}] obey the relation $\Delta \sim -R+\frac{1}{2}$; this indicates that the electron remains localised at the edge of the chain while the hole decays into the bulk.

\section{Experimental probe of shift exciton edge states}
\label{sec: experimental consequences}

\begin{figure}[t]
\centering
\includegraphics[width=0.48\textwidth]{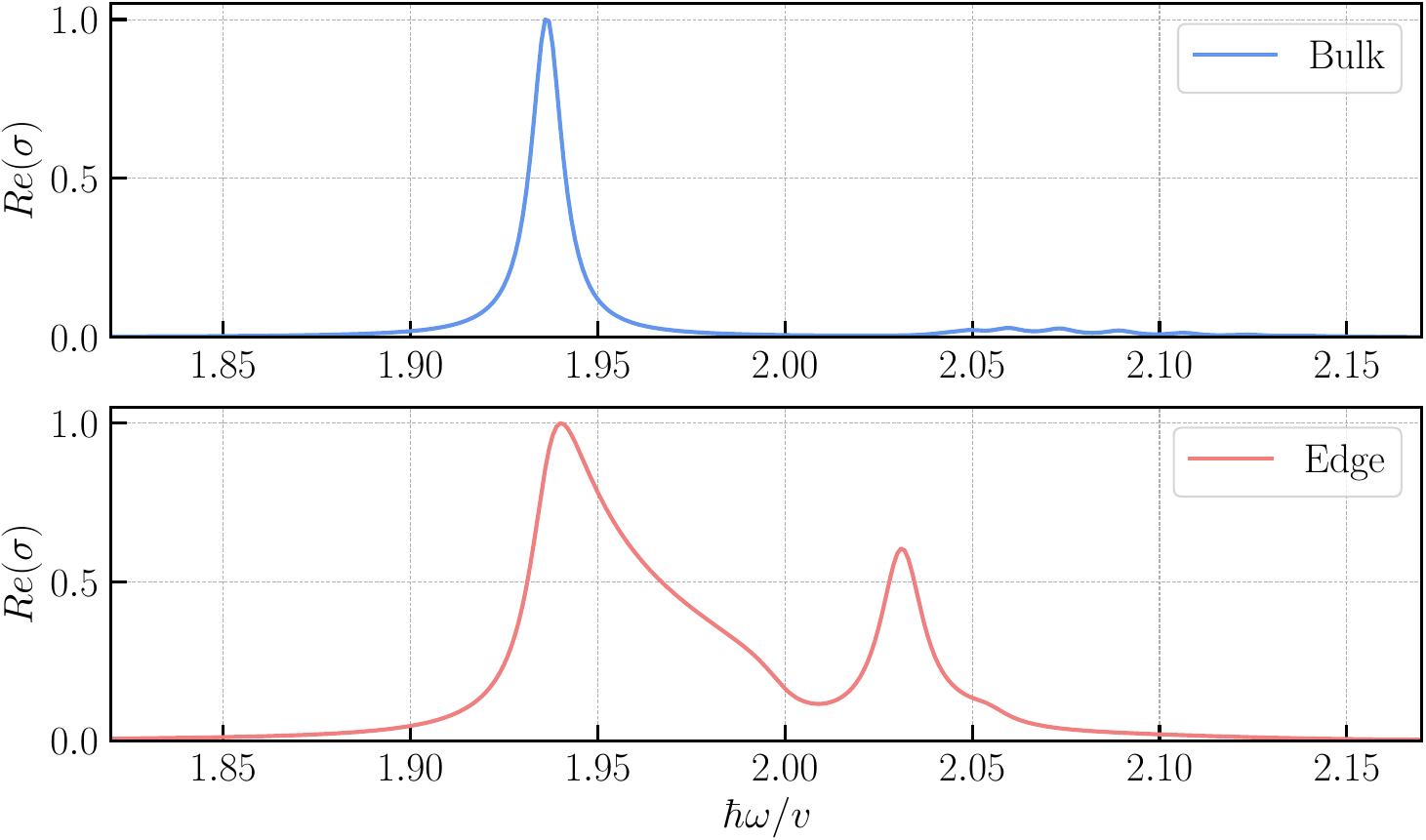}
\caption{Local optical conductivity in the bulk vs on the edge of the 1D SSH model [Eqs.~\eqref{eq: SSH Hamiltonian},~\eqref{eq: full gapping term}] with parameters $v=1.0,U=0.1, U'=0.1, w=0.03, V=0.01$.} 
\label{fig: optical conductivity}
\end{figure}

As proposed in Ref.~\onlinecite{topExcitonsMoire}, excitonic edge states can be directly probed via the local optical conductivity. The bulk optical conductivity can be calculated using the Kubo formula,
\begin{equation}
\sigma(\omega) = -\frac{1}{L} \sum_{n} \frac{|\bra{GS} \hat J \ket{\phi^{n}}|^2}{\hbar \omega -(E_n -E_0)+i \eta},
\end{equation}
where $\hat J = \sum_r \hat j(r)$, and $\hat j(r)$ is the local current operator~\cite{KuboFormula}. As discussed in Refs.~\onlinecite{localOpticCondExp, topExcitonsMoire}, a natural definition of the local optical conductivity, in terms of the local current operators, is
\begin{equation}
\sigma(\omega, r) = -\frac{1}{L} \sum_{n} \frac{\bra{GS} \hat J \ket{\phi^{n}}\bra{\phi^{n}} \hat j(r) \ket{GS}}{\hbar \omega -(E_n -E_0)+i \eta}.
\end{equation}
The local optical conductivity in the bulk is shown in Fig.~\ref{fig: optical conductivity}a and can be compared to that at the edge in Fig.~\ref{fig: optical conductivity}b. Both spectra show a contribution from the lowest energy band, but the edge excitons contribute a peak in the edge spectrum that is not seen in the bulk. The magnitude of this peak in the local optical conductivity decays exponentially into the bulk. From the normalisation of the $\ket{\phi^n}$ states, the local current matrix element for a given state can be seen to scale as $\sim\frac{1}{\sqrt{\xi}}$ where $\xi$ is the spatial extent of the state. The total current element integrates this quantity over the entire system and so scales as $\sim \frac{\xi}{\sqrt{\xi}}$. The product of the total and local current matrix elements therefore does not scale with the spatial extent of the states ($\xi$) for the bulk nor the edge states~\cite{topExcitonsMoire}. Hence, at the edge, neither peaks vanish in thermodynamic limit. The resulting change in the local optical conductivity at the edge can be probed using s-SNOM~\cite{localOpticCondExp} or spatially resolved absorption spectroscopy, and thus represents a clear measurable consequence of shift excitons.

\section{Discussion}
\label{sec: discussion}

Our work generalises the topological classification of TCIs to interaction-induced excitons in semiconductors, paving the way for future explorations into the interplay of topology and interaction-induced bound states in condensed matter systems. Firstly, investigating higher symmetry groups beyond $\mathcal{I}$ symmetry and applying topological quantum chemistry (TQC) principles to exciton band structures will uncover new topological phases of excitons. The composite nature of excitons also suggests the possibility of entirely novel topological properties not possible for (quasi-)electrons. Additionally, the condensation of shift excitons might lead to a new state of matter, termed a shift-excitonic insulator, with unique collective excitations and phase transitions. Another interesting direction would be to study the photovoltaic properties of shift excitons. Finally, our approach can be naturally generalised to other excitations, such as plasmons, magnons, polarons, magnon-magnon pairs, and triplons. 

\acknowledgments{We thank Gaurav Chaudhary, Julian May-Mann, and Ryan Barnett for inspiring discussions. We acknowledge support from the Imperial-TUM flagship partnership. JK acknowledges support from the Deutsche Forschungsgemeinschaft (DFG, German Research Foundation) under Germany’s Excellence Strategy–EXC– 2111–390814868, DFG grants No. KN1254/1-2, KN1254/2- 1, and TRR 360 - 492547816; as well as the Munich Quantum Valley, which is supported by the Bavarian state government with funds from the Hightech Agenda Bayern Plus.}
\bibliography{references}

\begin{thebibliography}{29}%
\makeatletter
\providecommand \@ifxundefined [1]{%
 \@ifx{#1\undefined}
}%
\providecommand \@ifnum [1]{%
 \ifnum #1\expandafter \@firstoftwo
 \else \expandafter \@secondoftwo
 \fi
}%
\providecommand \@ifx [1]{%
 \ifx #1\expandafter \@firstoftwo
 \else \expandafter \@secondoftwo
 \fi
}%
\providecommand \natexlab [1]{#1}%
\providecommand \enquote  [1]{``#1''}%
\providecommand \bibnamefont  [1]{#1}%
\providecommand \bibfnamefont [1]{#1}%
\providecommand \citenamefont [1]{#1}%
\providecommand \href@noop [0]{\@secondoftwo}%
\providecommand \href [0]{\begingroup \@sanitize@url \@href}%
\providecommand \@href[1]{\@@startlink{#1}\@@href}%
\providecommand \@@href[1]{\endgroup#1\@@endlink}%
\providecommand \@sanitize@url [0]{\catcode `\\12\catcode `\$12\catcode `\&12\catcode `\#12\catcode `\^12\catcode `\_12\catcode `\%12\relax}%
\providecommand \@@startlink[1]{}%
\providecommand \@@endlink[0]{}%
\providecommand \url  [0]{\begingroup\@sanitize@url \@url }%
\providecommand \@url [1]{\endgroup\@href {#1}{\urlprefix }}%
\providecommand \urlprefix  [0]{URL }%
\providecommand \Eprint [0]{\href }%
\providecommand \doibase [0]{https://doi.org/}%
\providecommand \selectlanguage [0]{\@gobble}%
\providecommand \bibinfo  [0]{\@secondoftwo}%
\providecommand \bibfield  [0]{\@secondoftwo}%
\providecommand \translation [1]{[#1]}%
\providecommand \BibitemOpen [0]{}%
\providecommand \bibitemStop [0]{}%
\providecommand \bibitemNoStop [0]{.\EOS\space}%
\providecommand \EOS [0]{\spacefactor3000\relax}%
\providecommand \BibitemShut  [1]{\csname bibitem#1\endcsname}%
\let\auto@bib@innerbib\@empty
\bibitem [{\citenamefont {Hasan}\ and\ \citenamefont {Kane}(2010)}]{HasanKaneRMP10}%
  \BibitemOpen
  \bibfield  {author} {\bibinfo {author} {\bibfnamefont {M.~Z.}\ \bibnamefont {Hasan}}\ and\ \bibinfo {author} {\bibfnamefont {C.~L.}\ \bibnamefont {Kane}},\ }\href {https://doi.org/10.1103/RevModPhys.82.3045} {\bibfield  {journal} {\bibinfo  {journal} {Rev. Mod. Phys.}\ }\textbf {\bibinfo {volume} {82}},\ \bibinfo {pages} {3045} (\bibinfo {year} {2010})}\BibitemShut {NoStop}%
\bibitem [{\citenamefont {Kane}\ and\ \citenamefont {Mele}(2005)}]{QSHEffectGrapheneMele2005}%
  \BibitemOpen
  \bibfield  {author} {\bibinfo {author} {\bibfnamefont {C.~L.}\ \bibnamefont {Kane}}\ and\ \bibinfo {author} {\bibfnamefont {E.~J.}\ \bibnamefont {Mele}},\ }\href {https://doi.org/10.1103/PhysRevLett.95.226801} {\bibfield  {journal} {\bibinfo  {journal} {Phys. Rev. Lett.}\ }\textbf {\bibinfo {volume} {95}},\ \bibinfo {pages} {226801} (\bibinfo {year} {2005})}\BibitemShut {NoStop}%
\bibitem [{\citenamefont {Bernevig}\ and\ \citenamefont {Zhang}(2006)}]{QSHEffectBernevig2006}%
  \BibitemOpen
  \bibfield  {author} {\bibinfo {author} {\bibfnamefont {B.~A.}\ \bibnamefont {Bernevig}}\ and\ \bibinfo {author} {\bibfnamefont {S.-C.}\ \bibnamefont {Zhang}},\ }\href {https://doi.org/10.1103/PhysRevLett.96.106802} {\bibfield  {journal} {\bibinfo  {journal} {Phys. Rev. Lett.}\ }\textbf {\bibinfo {volume} {96}},\ \bibinfo {pages} {106802} (\bibinfo {year} {2006})}\BibitemShut {NoStop}%
\bibitem [{\citenamefont {Fu}(2011)}]{FuTCI11}%
  \BibitemOpen
  \bibfield  {author} {\bibinfo {author} {\bibfnamefont {L.}~\bibnamefont {Fu}},\ }\href {https://doi.org/10.1103/PhysRevLett.106.106802} {\bibfield  {journal} {\bibinfo  {journal} {Phys. Rev. Lett.}\ }\textbf {\bibinfo {volume} {106}},\ \bibinfo {pages} {106802} (\bibinfo {year} {2011})}\BibitemShut {NoStop}%
\bibitem [{\citenamefont {Slager}\ \emph {et~al.}(2013)\citenamefont {Slager}, \citenamefont {Mesaros}, \citenamefont {Juri{\v c}i{\'c}},\ and\ \citenamefont {Zaanen}}]{topocrysInsulatorsRobertJan2013}%
  \BibitemOpen
  \bibfield  {author} {\bibinfo {author} {\bibfnamefont {R.-J.}\ \bibnamefont {Slager}}, \bibinfo {author} {\bibfnamefont {A.}~\bibnamefont {Mesaros}}, \bibinfo {author} {\bibfnamefont {V.}~\bibnamefont {Juri{\v c}i{\'c}}},\ and\ \bibinfo {author} {\bibfnamefont {J.}~\bibnamefont {Zaanen}},\ }\href {https://doi.org/10.1038/nphys2513} {\bibfield  {journal} {\bibinfo  {journal} {Nature Physics}\ }\textbf {\bibinfo {volume} {9}},\ \bibinfo {pages} {98} (\bibinfo {year} {2013})}\BibitemShut {NoStop}%
\bibitem [{\citenamefont {Kruthoff}\ \emph {et~al.}(2017)\citenamefont {Kruthoff}, \citenamefont {de~Boer}, \citenamefont {van Wezel}, \citenamefont {Kane},\ and\ \citenamefont {Slager}}]{RobertJanPRX17}%
  \BibitemOpen
  \bibfield  {author} {\bibinfo {author} {\bibfnamefont {J.}~\bibnamefont {Kruthoff}}, \bibinfo {author} {\bibfnamefont {J.}~\bibnamefont {de~Boer}}, \bibinfo {author} {\bibfnamefont {J.}~\bibnamefont {van Wezel}}, \bibinfo {author} {\bibfnamefont {C.~L.}\ \bibnamefont {Kane}},\ and\ \bibinfo {author} {\bibfnamefont {R.-J.}\ \bibnamefont {Slager}},\ }\href {https://doi.org/10.1103/PhysRevX.7.041069} {\bibfield  {journal} {\bibinfo  {journal} {Phys. Rev. X}\ }\textbf {\bibinfo {volume} {7}},\ \bibinfo {pages} {041069} (\bibinfo {year} {2017})}\BibitemShut {NoStop}%
\bibitem [{\citenamefont {Po}\ \emph {et~al.}(2017)\citenamefont {Po}, \citenamefont {Vishwanath},\ and\ \citenamefont {Watanabe}}]{Ashvin230SINatComm17}%
  \BibitemOpen
  \bibfield  {author} {\bibinfo {author} {\bibfnamefont {H.~C.}\ \bibnamefont {Po}}, \bibinfo {author} {\bibfnamefont {A.}~\bibnamefont {Vishwanath}},\ and\ \bibinfo {author} {\bibfnamefont {H.}~\bibnamefont {Watanabe}},\ }\href {https://doi.org/10.1038/s41467-017-00133-2} {\bibfield  {journal} {\bibinfo  {journal} {Nature Communications}\ }\textbf {\bibinfo {volume} {8}},\ \bibinfo {pages} {50} (\bibinfo {year} {2017})}\BibitemShut {NoStop}%
\bibitem [{\citenamefont {Bradlyn}\ \emph {et~al.}(2017)\citenamefont {Bradlyn}, \citenamefont {Elcoro}, \citenamefont {Cano}, \citenamefont {Vergniory}, \citenamefont {Wang}, \citenamefont {Felser}, \citenamefont {Aroyo},\ and\ \citenamefont {Bernevig}}]{AndreiTQC17}%
  \BibitemOpen
  \bibfield  {author} {\bibinfo {author} {\bibfnamefont {B.}~\bibnamefont {Bradlyn}}, \bibinfo {author} {\bibfnamefont {L.}~\bibnamefont {Elcoro}}, \bibinfo {author} {\bibfnamefont {J.}~\bibnamefont {Cano}}, \bibinfo {author} {\bibfnamefont {M.~G.}\ \bibnamefont {Vergniory}}, \bibinfo {author} {\bibfnamefont {Z.}~\bibnamefont {Wang}}, \bibinfo {author} {\bibfnamefont {C.}~\bibnamefont {Felser}}, \bibinfo {author} {\bibfnamefont {M.~I.}\ \bibnamefont {Aroyo}},\ and\ \bibinfo {author} {\bibfnamefont {B.~A.}\ \bibnamefont {Bernevig}},\ }\href {https://doi.org/10.1038/nature23268} {\bibfield  {journal} {\bibinfo  {journal} {Nature}\ }\textbf {\bibinfo {volume} {547}},\ \bibinfo {pages} {298} (\bibinfo {year} {2017})}\BibitemShut {NoStop}%
\bibitem [{\citenamefont {Song}\ \emph {et~al.}(2019)\citenamefont {Song}, \citenamefont {Huang}, \citenamefont {Qi}, \citenamefont {Fang},\ and\ \citenamefont {Hermele}}]{HermeleTopoCrystalSciAdv19}%
  \BibitemOpen
  \bibfield  {author} {\bibinfo {author} {\bibfnamefont {Z.}~\bibnamefont {Song}}, \bibinfo {author} {\bibfnamefont {S.-J.}\ \bibnamefont {Huang}}, \bibinfo {author} {\bibfnamefont {Y.}~\bibnamefont {Qi}}, \bibinfo {author} {\bibfnamefont {C.}~\bibnamefont {Fang}},\ and\ \bibinfo {author} {\bibfnamefont {M.}~\bibnamefont {Hermele}},\ }\href {https://doi.org/10.1126/sciadv.aax2007} {\bibfield  {journal} {\bibinfo  {journal} {Science Advances}\ }\textbf {\bibinfo {volume} {5}},\ \bibinfo {pages} {eaax2007} (\bibinfo {year} {2019})},\ \Eprint {https://arxiv.org/abs/https://www.science.org/doi/pdf/10.1126/sciadv.aax2007} {https://www.science.org/doi/pdf/10.1126/sciadv.aax2007} \BibitemShut {NoStop}%
\bibitem [{\citenamefont {Shiozaki}\ \emph {et~al.}(2022)\citenamefont {Shiozaki}, \citenamefont {Sato},\ and\ \citenamefont {Gomi}}]{ShiozakiSpectralSequencePRB22}%
  \BibitemOpen
  \bibfield  {author} {\bibinfo {author} {\bibfnamefont {K.}~\bibnamefont {Shiozaki}}, \bibinfo {author} {\bibfnamefont {M.}~\bibnamefont {Sato}},\ and\ \bibinfo {author} {\bibfnamefont {K.}~\bibnamefont {Gomi}},\ }\href {https://doi.org/10.1103/PhysRevB.106.165103} {\bibfield  {journal} {\bibinfo  {journal} {Phys. Rev. B}\ }\textbf {\bibinfo {volume} {106}},\ \bibinfo {pages} {165103} (\bibinfo {year} {2022})}\BibitemShut {NoStop}%
\bibitem [{\citenamefont {Soldini}\ \emph {et~al.}(2023)\citenamefont {Soldini}, \citenamefont {Astrakhantsev}, \citenamefont {Iraola}, \citenamefont {Tiwari}, \citenamefont {Fischer}, \citenamefont {Valent\'{\i}}, \citenamefont {Vergniory}, \citenamefont {Wagner},\ and\ \citenamefont {Neupert}}]{MartinaInteractingTQC23}%
  \BibitemOpen
  \bibfield  {author} {\bibinfo {author} {\bibfnamefont {M.~O.}\ \bibnamefont {Soldini}}, \bibinfo {author} {\bibfnamefont {N.}~\bibnamefont {Astrakhantsev}}, \bibinfo {author} {\bibfnamefont {M.}~\bibnamefont {Iraola}}, \bibinfo {author} {\bibfnamefont {A.}~\bibnamefont {Tiwari}}, \bibinfo {author} {\bibfnamefont {M.~H.}\ \bibnamefont {Fischer}}, \bibinfo {author} {\bibfnamefont {R.}~\bibnamefont {Valent\'{\i}}}, \bibinfo {author} {\bibfnamefont {M.~G.}\ \bibnamefont {Vergniory}}, \bibinfo {author} {\bibfnamefont {G.}~\bibnamefont {Wagner}},\ and\ \bibinfo {author} {\bibfnamefont {T.}~\bibnamefont {Neupert}},\ }\href {https://doi.org/10.1103/PhysRevB.107.245145} {\bibfield  {journal} {\bibinfo  {journal} {Phys. Rev. B}\ }\textbf {\bibinfo {volume} {107}},\ \bibinfo {pages} {245145} (\bibinfo {year} {2023})}\BibitemShut {NoStop}%
\bibitem [{\citenamefont {Shiozaki}\ \emph {et~al.}(2023)\citenamefont {Shiozaki}, \citenamefont {Xiong},\ and\ \citenamefont {Gomi}}]{ShiozakiSPTClassification23}%
  \BibitemOpen
  \bibfield  {author} {\bibinfo {author} {\bibfnamefont {K.}~\bibnamefont {Shiozaki}}, \bibinfo {author} {\bibfnamefont {C.~Z.}\ \bibnamefont {Xiong}},\ and\ \bibinfo {author} {\bibfnamefont {K.}~\bibnamefont {Gomi}},\ }\href {https://doi.org/10.1093/ptep/ptad086} {\bibfield  {journal} {\bibinfo  {journal} {Progress of Theoretical and Experimental Physics}\ }\textbf {\bibinfo {volume} {2023}},\ \bibinfo {pages} {083I01} (\bibinfo {year} {2023})},\ \Eprint {https://arxiv.org/abs/https://academic.oup.com/ptep/article-pdf/2023/8/083I01/51118278/ptad086.pdf} {https://academic.oup.com/ptep/article-pdf/2023/8/083I01/51118278/ptad086.pdf} \BibitemShut {NoStop}%
\bibitem [{\citenamefont {Manjunath}\ \emph {et~al.}(2024)\citenamefont {Manjunath}, \citenamefont {Calvera},\ and\ \citenamefont {Barkeshli}}]{NarenManjunathTCIClassification24}%
  \BibitemOpen
  \bibfield  {author} {\bibinfo {author} {\bibfnamefont {N.}~\bibnamefont {Manjunath}}, \bibinfo {author} {\bibfnamefont {V.}~\bibnamefont {Calvera}},\ and\ \bibinfo {author} {\bibfnamefont {M.}~\bibnamefont {Barkeshli}},\ }\href {https://doi.org/10.1103/PhysRevB.109.035168} {\bibfield  {journal} {\bibinfo  {journal} {Phys. Rev. B}\ }\textbf {\bibinfo {volume} {109}},\ \bibinfo {pages} {035168} (\bibinfo {year} {2024})}\BibitemShut {NoStop}%
\bibitem [{\citenamefont {Herzog-Arbeitman}\ \emph {et~al.}(2024)\citenamefont {Herzog-Arbeitman}, \citenamefont {Bernevig},\ and\ \citenamefont {Song}}]{JonahInteractingTQCNatComm24}%
  \BibitemOpen
  \bibfield  {author} {\bibinfo {author} {\bibfnamefont {J.}~\bibnamefont {Herzog-Arbeitman}}, \bibinfo {author} {\bibfnamefont {B.~A.}\ \bibnamefont {Bernevig}},\ and\ \bibinfo {author} {\bibfnamefont {Z.-D.}\ \bibnamefont {Song}},\ }\href {https://doi.org/10.1038/s41467-024-45395-9} {\bibfield  {journal} {\bibinfo  {journal} {Nature Communications}\ }\textbf {\bibinfo {volume} {15}},\ \bibinfo {pages} {1171} (\bibinfo {year} {2024})}\BibitemShut {NoStop}%
\bibitem [{\citenamefont {Chen}\ and\ \citenamefont {Shindou}(2017)}]{ChenPhysRevB2017}%
  \BibitemOpen
  \bibfield  {author} {\bibinfo {author} {\bibfnamefont {K.}~\bibnamefont {Chen}}\ and\ \bibinfo {author} {\bibfnamefont {R.}~\bibnamefont {Shindou}},\ }\href {https://doi.org/10.1103/PhysRevB.96.161101} {\bibfield  {journal} {\bibinfo  {journal} {Phys. Rev. B}\ }\textbf {\bibinfo {volume} {96}},\ \bibinfo {pages} {161101} (\bibinfo {year} {2017})}\BibitemShut {NoStop}%
\bibitem [{\citenamefont {Gong}\ \emph {et~al.}(2017)\citenamefont {Gong}, \citenamefont {Luo}, \citenamefont {Jiang},\ and\ \citenamefont {Fu}}]{FuChernExcitons}%
  \BibitemOpen
  \bibfield  {author} {\bibinfo {author} {\bibfnamefont {Z.~R.}\ \bibnamefont {Gong}}, \bibinfo {author} {\bibfnamefont {W.~Z.}\ \bibnamefont {Luo}}, \bibinfo {author} {\bibfnamefont {Z.~F.}\ \bibnamefont {Jiang}},\ and\ \bibinfo {author} {\bibfnamefont {H.~C.}\ \bibnamefont {Fu}},\ }\href {https://doi.org/10.1038/srep42390} {\bibfield  {journal} {\bibinfo  {journal} {Scientific Reports}\ }\textbf {\bibinfo {volume} {7}},\ \bibinfo {pages} {42390} (\bibinfo {year} {2017})}\BibitemShut {NoStop}%
\bibitem [{\citenamefont {Xie}\ \emph {et~al.}(2024)\citenamefont {Xie}, \citenamefont {Ghaemi}, \citenamefont {Mitrano},\ and\ \citenamefont {Uchoa}}]{xie2024theory}%
  \BibitemOpen
  \bibfield  {author} {\bibinfo {author} {\bibfnamefont {H.-Y.}\ \bibnamefont {Xie}}, \bibinfo {author} {\bibfnamefont {P.}~\bibnamefont {Ghaemi}}, \bibinfo {author} {\bibfnamefont {M.}~\bibnamefont {Mitrano}},\ and\ \bibinfo {author} {\bibfnamefont {B.}~\bibnamefont {Uchoa}},\ }\href@noop {} {\bibinfo {title} {Theory of topological exciton insulators and condensates in flat chern bands}} (\bibinfo {year} {2024}),\ \Eprint {https://arxiv.org/abs/2311.04970} {arXiv:2311.04970 [cond-mat.mes-hall]} \BibitemShut {NoStop}%
\bibitem [{\citenamefont {Kwan}\ \emph {et~al.}(2021)\citenamefont {Kwan}, \citenamefont {Hu}, \citenamefont {Simon},\ and\ \citenamefont {Parameswaran}}]{ParamesaranPhysRevLet2021}%
  \BibitemOpen
  \bibfield  {author} {\bibinfo {author} {\bibfnamefont {Y.~H.}\ \bibnamefont {Kwan}}, \bibinfo {author} {\bibfnamefont {Y.}~\bibnamefont {Hu}}, \bibinfo {author} {\bibfnamefont {S.~H.}\ \bibnamefont {Simon}},\ and\ \bibinfo {author} {\bibfnamefont {S.~A.}\ \bibnamefont {Parameswaran}},\ }\href {https://doi.org/10.1103/PhysRevLett.126.137601} {\bibfield  {journal} {\bibinfo  {journal} {Phys. Rev. Lett.}\ }\textbf {\bibinfo {volume} {126}},\ \bibinfo {pages} {137601} (\bibinfo {year} {2021})}\BibitemShut {NoStop}%
\bibitem [{\citenamefont {Blason}\ and\ \citenamefont {Fabrizio}(2020)}]{QSHInsulatorExcitonsFabrizio}%
  \BibitemOpen
  \bibfield  {author} {\bibinfo {author} {\bibfnamefont {A.}~\bibnamefont {Blason}}\ and\ \bibinfo {author} {\bibfnamefont {M.}~\bibnamefont {Fabrizio}},\ }\href {https://doi.org/10.1103/PhysRevB.102.035146} {\bibfield  {journal} {\bibinfo  {journal} {Phys. Rev. B}\ }\textbf {\bibinfo {volume} {102}},\ \bibinfo {pages} {035146} (\bibinfo {year} {2020})}\BibitemShut {NoStop}%
\bibitem [{\citenamefont {Haber}\ \emph {et~al.}(2023)\citenamefont {Haber}, \citenamefont {Qiu}, \citenamefont {da~Jornada},\ and\ \citenamefont {Neaton}}]{NeatonExcitonMLWFsPRB23}%
  \BibitemOpen
  \bibfield  {author} {\bibinfo {author} {\bibfnamefont {J.~B.}\ \bibnamefont {Haber}}, \bibinfo {author} {\bibfnamefont {D.~Y.}\ \bibnamefont {Qiu}}, \bibinfo {author} {\bibfnamefont {F.~H.}\ \bibnamefont {da~Jornada}},\ and\ \bibinfo {author} {\bibfnamefont {J.~B.}\ \bibnamefont {Neaton}},\ }\href {https://doi.org/10.1103/PhysRevB.108.125118} {\bibfield  {journal} {\bibinfo  {journal} {Phys. Rev. B}\ }\textbf {\bibinfo {volume} {108}},\ \bibinfo {pages} {125118} (\bibinfo {year} {2023})}\BibitemShut {NoStop}%
\bibitem [{\citenamefont {Su}\ \emph {et~al.}(1979)\citenamefont {Su}, \citenamefont {Schrieffer},\ and\ \citenamefont {Heeger}}]{SSHPRL79}%
  \BibitemOpen
  \bibfield  {author} {\bibinfo {author} {\bibfnamefont {W.~P.}\ \bibnamefont {Su}}, \bibinfo {author} {\bibfnamefont {J.~R.}\ \bibnamefont {Schrieffer}},\ and\ \bibinfo {author} {\bibfnamefont {A.~J.}\ \bibnamefont {Heeger}},\ }\href {https://doi.org/10.1103/PhysRevLett.42.1698} {\bibfield  {journal} {\bibinfo  {journal} {Phys. Rev. Lett.}\ }\textbf {\bibinfo {volume} {42}},\ \bibinfo {pages} {1698} (\bibinfo {year} {1979})}\BibitemShut {NoStop}%
\bibitem [{\citenamefont {Luo}\ \emph {et~al.}(2020)\citenamefont {Luo}, \citenamefont {Engelke}, \citenamefont {Mattheakis}, \citenamefont {Tamagnone}, \citenamefont {Carr}, \citenamefont {Watanabe}, \citenamefont {Taniguchi}, \citenamefont {Kaxiras}, \citenamefont {Kim},\ and\ \citenamefont {Wilson}}]{localOpticCondExp}%
  \BibitemOpen
  \bibfield  {author} {\bibinfo {author} {\bibfnamefont {Y.}~\bibnamefont {Luo}}, \bibinfo {author} {\bibfnamefont {R.}~\bibnamefont {Engelke}}, \bibinfo {author} {\bibfnamefont {M.}~\bibnamefont {Mattheakis}}, \bibinfo {author} {\bibfnamefont {M.}~\bibnamefont {Tamagnone}}, \bibinfo {author} {\bibfnamefont {S.}~\bibnamefont {Carr}}, \bibinfo {author} {\bibfnamefont {K.}~\bibnamefont {Watanabe}}, \bibinfo {author} {\bibfnamefont {T.}~\bibnamefont {Taniguchi}}, \bibinfo {author} {\bibfnamefont {E.}~\bibnamefont {Kaxiras}}, \bibinfo {author} {\bibfnamefont {P.}~\bibnamefont {Kim}},\ and\ \bibinfo {author} {\bibfnamefont {W.~L.}\ \bibnamefont {Wilson}},\ }\href@noop {} {\bibfield  {journal} {\bibinfo  {journal} {Nature Communications}\ }\textbf {\bibinfo {volume} {11}},\ \bibinfo {pages} {4209} (\bibinfo {year} {2020})}\BibitemShut {NoStop}%
\bibitem [{\citenamefont {Wu}\ \emph {et~al.}(2017)\citenamefont {Wu}, \citenamefont {Lovorn},\ and\ \citenamefont {MacDonald}}]{topExcitonsMoire}%
  \BibitemOpen
  \bibfield  {author} {\bibinfo {author} {\bibfnamefont {F.}~\bibnamefont {Wu}}, \bibinfo {author} {\bibfnamefont {T.}~\bibnamefont {Lovorn}},\ and\ \bibinfo {author} {\bibfnamefont {A.~H.}\ \bibnamefont {MacDonald}},\ }\href {https://doi.org/10.1103/PhysRevLett.118.147401} {\bibfield  {journal} {\bibinfo  {journal} {Phys. Rev. Lett.}\ }\textbf {\bibinfo {volume} {118}},\ \bibinfo {pages} {147401} (\bibinfo {year} {2017})}\BibitemShut {NoStop}%
\bibitem [{\citenamefont {Alexandradinata}\ \emph {et~al.}(2014)\citenamefont {Alexandradinata}, \citenamefont {Dai},\ and\ \citenamefont {Bernevig}}]{AlexandradinataWilsonPRB14}%
  \BibitemOpen
  \bibfield  {author} {\bibinfo {author} {\bibfnamefont {A.}~\bibnamefont {Alexandradinata}}, \bibinfo {author} {\bibfnamefont {X.}~\bibnamefont {Dai}},\ and\ \bibinfo {author} {\bibfnamefont {B.~A.}\ \bibnamefont {Bernevig}},\ }\href {https://doi.org/10.1103/PhysRevB.89.155114} {\bibfield  {journal} {\bibinfo  {journal} {Phys. Rev. B}\ }\textbf {\bibinfo {volume} {89}},\ \bibinfo {pages} {155114} (\bibinfo {year} {2014})}\BibitemShut {NoStop}%
\bibitem [{\citenamefont {Marzari}\ and\ \citenamefont {Vanderbilt}(1997)}]{MarzariVanderbiltSmoothGaugePRB97}%
  \BibitemOpen
  \bibfield  {author} {\bibinfo {author} {\bibfnamefont {N.}~\bibnamefont {Marzari}}\ and\ \bibinfo {author} {\bibfnamefont {D.}~\bibnamefont {Vanderbilt}},\ }\href {https://doi.org/10.1103/PhysRevB.56.12847} {\bibfield  {journal} {\bibinfo  {journal} {Phys. Rev. B}\ }\textbf {\bibinfo {volume} {56}},\ \bibinfo {pages} {12847} (\bibinfo {year} {1997})}\BibitemShut {NoStop}%
\bibitem [{\citenamefont {Benalcazar}\ \emph {et~al.}(2019)\citenamefont {Benalcazar}, \citenamefont {Li},\ and\ \citenamefont {Hughes}}]{BenalcazarCnCornerStatesPRB19}%
  \BibitemOpen
  \bibfield  {author} {\bibinfo {author} {\bibfnamefont {W.~A.}\ \bibnamefont {Benalcazar}}, \bibinfo {author} {\bibfnamefont {T.}~\bibnamefont {Li}},\ and\ \bibinfo {author} {\bibfnamefont {T.~L.}\ \bibnamefont {Hughes}},\ }\href {https://doi.org/10.1103/PhysRevB.99.245151} {\bibfield  {journal} {\bibinfo  {journal} {Phys. Rev. B}\ }\textbf {\bibinfo {volume} {99}},\ \bibinfo {pages} {245151} (\bibinfo {year} {2019})}\BibitemShut {NoStop}%
\bibitem [{\citenamefont {Asb{\'o}th}\ \emph {et~al.}(2016)\citenamefont {Asb{\'o}th}, \citenamefont {Oroszl{\'a}ny},\ and\ \citenamefont {P{\'a}lyi}}]{AsbothBook16}%
  \BibitemOpen
  \bibfield  {author} {\bibinfo {author} {\bibfnamefont {J.~K.}\ \bibnamefont {Asb{\'o}th}}, \bibinfo {author} {\bibfnamefont {L.}~\bibnamefont {Oroszl{\'a}ny}},\ and\ \bibinfo {author} {\bibfnamefont {A.}~\bibnamefont {P{\'a}lyi}},\ }\bibinfo {title} {The su-schrieffer-heeger (ssh) model},\ in\ \href {https://doi.org/10.1007/978-3-319-25607-8{\_}1} {\emph {\bibinfo {booktitle} {A Short Course on Topological Insulators: Band Structure and Edge States in One and Two Dimensions}}},\ \bibinfo {editor} {edited by\ \bibinfo {editor} {\bibfnamefont {J.~K.}\ \bibnamefont {Asb{\'o}th}}, \bibinfo {editor} {\bibfnamefont {L.}~\bibnamefont {Oroszl{\'a}ny}},\ and\ \bibinfo {editor} {\bibfnamefont {A.}~\bibnamefont {P{\'a}lyi}}}\ (\bibinfo  {publisher} {Springer International Publishing},\ \bibinfo {address} {Cham},\ \bibinfo {year} {2016})\ pp.\ \bibinfo {pages} {1--22}\BibitemShut {NoStop}%
\bibitem [{\citenamefont {Schindler}\ and\ \citenamefont {Bernevig}(2021)}]{NoncompactPRB21}%
  \BibitemOpen
  \bibfield  {author} {\bibinfo {author} {\bibfnamefont {F.}~\bibnamefont {Schindler}}\ and\ \bibinfo {author} {\bibfnamefont {B.~A.}\ \bibnamefont {Bernevig}},\ }\href {https://doi.org/10.1103/PhysRevB.104.L201114} {\bibfield  {journal} {\bibinfo  {journal} {Phys. Rev. B}\ }\textbf {\bibinfo {volume} {104}},\ \bibinfo {pages} {L201114} (\bibinfo {year} {2021})}\BibitemShut {NoStop}%
\bibitem [{\citenamefont {Kubo}(1957)}]{KuboFormula}%
  \BibitemOpen
  \bibfield  {author} {\bibinfo {author} {\bibfnamefont {R.}~\bibnamefont {Kubo}},\ }\href {https://doi.org/10.1143/JPSJ.12.570} {\bibfield  {journal} {\bibinfo  {journal} {Journal of the Physical Society of Japan}\ }\textbf {\bibinfo {volume} {12}},\ \bibinfo {pages} {570} (\bibinfo {year} {1957})},\ \Eprint {https://arxiv.org/abs/https://doi.org/10.1143/JPSJ.12.570} {https://doi.org/10.1143/JPSJ.12.570} \BibitemShut {NoStop}%
\end{thebibliography}%

\appendix

\section{Definition of Shift Excitons}
\label{apdx:ShiftDefinition}
The main text introduces shift excitons in trivial bands; in this appendix we generalise the notion of shift excitons to nontrivial bands \emph{i.e.} those where the exciton Wannier centre is no longer at the origin of the unit cell. We define the shift using the projected position operator for electrons in the empty band and holes in the filled band. The microscopic electron position operator is
\begin{equation}
\hat X = \sum_{R, i} R c^\dagger_{R, i} c_{R, i}.
\end{equation}
This can be projected into a single band. The projected position operator for electrons in a given band labelled $\alpha$ is,
\begin{equation}
\hat x^{(\mathrm{e})}_{\alpha} = \sum_{R} (R+x_\alpha) c^\dagger_{R, \alpha} c_{R, \alpha},
\end{equation}
where $x_\alpha$ is the electron Wannier centre within the unit cell. This is $x_\alpha=0$ for trivial bands. Similarly, the projected position operator for the holes in a given band $\alpha$ is,
\begin{equation}
\hat x^{(\mathrm{h})}_{\alpha} = \sum_{R} (R+x_\alpha) c_{R, \alpha} c^\dagger_{R, \alpha}.
\end{equation}
The shift in the exciton Wannier state corresponds to a shift in electron and hole positions within the exciton Wannier state. Therefore we calculate expectation value of the band projected operators in the exciton Wannier states for both electrons in the empty band and holes in the occupied band. For the exciton Wannier state $\ket{W^{R'}}$ centered at $R'$ we calculate,
\begin{equation}
\begin{aligned}
\bra{W^{R'=0}} &\hat x_{\mathrm{emp}}^{(\mathrm{e})}\ket{W^{R'=0}} \\&= \bra{GS} \bigg(\sum_{\tilde R, \tilde \Delta} c^\dagger_{\tilde R-\tilde \Delta, \mathrm{occ}} c_{\tilde R, \mathrm{emp}} (W^{\tilde R}_{\tilde \Delta})^* \bigg)\cdot \\&\bigg(\sum_{R}(R+x_{\mathrm{emp}})c^\dagger_{R, \mathrm{emp}}c_{R, \mathrm{emp}}\bigg)\cdot \\&\bigg(\sum_{\tilde R', \tilde \Delta'} W^{\tilde R'}_{\tilde \Delta'} c^\dagger_{\tilde R', \mathrm{emp}} c_{\tilde R'-\tilde \Delta', \mathrm{occ}}\bigg) \ket{GS}.
\end{aligned}
\end{equation}
This expression can be simplified to give,
\begin{equation}
\begin{aligned}
&\bra{W^{R'=0}} \hat x_{\mathrm{emp}}^{(\mathrm{e})}\ket{W^{R'=0}} \\
&= \sum_{R, \tilde R, \tilde\Delta, \tilde R', \tilde \Delta'} (R+x_{\mathrm{emp}}) (W^{\tilde R'}_{\tilde \Delta'})^* (W^{\tilde R}_{\tilde \Delta})
\delta_{\tilde R', R}\delta_{\tilde R, R} \delta_{\tilde \Delta, \tilde \Delta'}\\
&= \sum_{R, \Delta} (R+x_{\mathrm{emp}})|W^{\tilde R'}_{\tilde \Delta'}|^2 \\
&= \sum_{R, \Delta} \big(|W^{R}_{\Delta}|^2 R\big) + x_{\mathrm{emp}}.
\end{aligned}
\end{equation}
Similarly, the expectation value of the hole position in the occupied band is,
\begin{equation}
\begin{aligned}
\bra{W^{R'=0}} \hat x_{\mathrm{occ}}^{(\mathrm{h})}\ket{W^{R'=0}}  &= \sum_{R, \Delta} |W^{R}_{\Delta}|^2 \big(R-\Delta + x_{\mathrm{occ}}\big)\\
&= \sum_{R, \Delta} \big(|W^{R}_{\Delta}|^2 R\big) + x_{\mathrm{occ}}.
\end{aligned}
\end{equation}
This was simplified using $\sum_{R, \Delta} \big(|W^{R}_{\Delta}|^2 \Delta\big) = 0$ for both trivial and nontrivial exciton bands. 

The exciton shift ($s_{\mathrm{exc}}$) is therefore defined as,
\begin{equation}
s_{\mathrm{exc}} = \braket{W^{0}| \hat{x}^{(\mathrm{e})}_\mathrm{emp} | W^{0}} - x_\mathrm{emp}
= \braket{W^{0}| \hat{x}^{(\mathrm{h})}_\mathrm{occ} | W^{0}} - x_\mathrm{occ}.
\end{equation}
It follows that the shift ($s_{exc}$) of the exciton Wannier state represents the shift in the electron and hole positions (within the exciton Wannier state) with respect to the noninteracting electron and hole Wannier centers respectively. Regardless of the noninteracting electron Wannier centers, therefore the exciton Wannier state shift is defined as,
\begin{equation}
s_{exc} = \sum_{R, \Delta} |W^{R}_{\Delta}|^2 R.
\end{equation}

\section{Theory for Generic Hopping Terms and Interactions}
We here derive the most general exciton Hamiltonian that results from perturbing the dimerised-limit SSH model with arbitrary hopping terms and interactions.
\label{apdx:GeneralTheory}
\subsection{Generic Hopping Terms}
We consider a 1D system with two sites per unit cell in periodic boundary conditions (PBC). We set the intercell hopping $v$ to unity and treat further hoppings and interactions perturbatively. The main text treated the results for just an additional nearest neighbour hopping $w$. Here we present the effective exciton Hamiltonian for generic hoppings \emph{e.g.},
\begin{equation}
\hat H' = \sum_{\substack{R, R'\\ i, j \in \{A,B\}}} t_{R, R', i, j}c^\dagger_{R, i} c_{R', j}.
\label{eq:hoppingTerm}
\end{equation}
As introduced in the main text, we treat these hoppings perturbatively on top of the noninteracting ground state of the dimerised SSH model (i.e., the ground state at $v = 1, U = U' = 0$ and all $t_{R, R', i, j} = 0$).
The hopping term in Eq.~\eqref{eq:hoppingTerm} can be rewritten in terms of the Wannier states of the dimerised limit model (e.g. \ref{eq:electronWannierStatesSSH}) hence, 
\begin{equation}
c^\dagger_{R, \alpha} = \sum_{i\in \{A, B\}} M_{i, \alpha} c^\dagger_{R, i},
\end{equation}
where 
\begin{equation}
M_{i\alpha} = \frac{1}{\sqrt{2}}\begin{pmatrix}
    1 & 1\\
1 & -1
\end{pmatrix}.   
\end{equation}
The indices $i, \alpha$ give the first row or column for $i=A$ or $\alpha=+$  and the second row and column for $i=B$ or $\alpha=-$ .
The hopping Hamiltonian in equation \ref{eq:hoppingTerm} can therefore be rewritten in this Wannier basis as,
\begin{equation}
\hat H' = \sum_{\substack{R, R'\\ \alpha, \beta \in \{+,-\}}}t_{R, R', \alpha, \beta}c^\dagger_{R, \alpha} c_{R', \beta},
\end{equation}
where $t_{R, R', \alpha, \beta} = \sum_{i,j} t_{R,R', i, j} M_{i, \alpha} M^*_{j,\beta}$.

We apply the Hamiltonian $\hat H'$ to our variational exciton basis state $c^\dagger_{R+\Delta, -} c_{R,+} \ket{GS}$. This gives
\begin{equation}
\begin{aligned}
&\sum_{\substack{\tilde R, \tilde R'\\ \alpha, \beta \in \{+, -\}}} t_{\tilde R, \tilde R' , \alpha , \beta} c^\dagger_{\tilde R, \alpha} c_{\tilde R', \beta} c^\dagger_{R+\Delta, -}c_{R, +} \ket{GS}\\
&= \sum_{\tilde R} t_{\tilde R, R+\Delta, --} c^\dagger_{\tilde R, -}c_{R, +} \ket{GS}\\ &- \sum_{\tilde R} t_{ R, \tilde R, ++} c^\dagger_{R+\Delta, -}c_{\tilde R, +} \ket{GS}\\&+(t_{R+\Delta, R+\Delta, --} + \sum_{\tilde R \neq R} t_{\tilde R, \tilde R, ++}) c^\dagger_{R+\Delta, -} c_{R,+} \ket{GS}\\
&+ (\text{terms outside of variational basis}).
\label{eq:hoppingtermCalculation}
\end{aligned}
\end{equation}
In addition, the ground state energy shifts due to $\hat H'$, the ground state expectation value is
\begin{align}
\bra{GS} \hat H' \ket{GS} &= \sum_{R, R', \alpha, \beta} t_{R, R' \alpha \beta} \bra{GS} c^\dagger_{R,\alpha} c_{R', \beta} \ket{GS}\\
&= \sum_{\tilde R} t_{\tilde R, \tilde R, ++}.
\label{eq:GSExpectationVal}
\end{align}
We use the Eq.~\eqref{eq:hoppingtermCalculation} and Eq.~\eqref{eq:GSExpectationVal} to calculate the matrix elements in the variational basis (minus the ground state expectation value),
\begin{equation}
\begin{aligned}
&H'_{R',R,\Delta',\Delta} \equiv \\&\bra{GS} c^\dagger_{R', +}c_{R'+\Delta', -} \hat H' c^\dagger_{R+\Delta, -} c_{R, +} \ket{GS}\\ &- \bra{GS} \hat H' \ket{GS} \\
&= \delta_{\Delta', \Delta} \delta_{R', R} (t_{00--}-t_{00++})\\ &+ \delta_{R', R} t_{\Delta'\Delta--}- \delta_{\Delta', \Delta+R-R'} t_{RR'++}.
\label{eq:EffExcHamRDeltaSpace}
\end{aligned}
\end{equation}
Here we use the translational symmetry $t_{R, R',\alpha\beta} = t_{R+x, R'+x, \alpha\beta}$ to simplify the expression. 

As explained in the main text, this resulting Hamiltonian is translationally invariant so that $H'_{R',R,\Delta',\Delta} = H'_{R'-R, 0, \Delta',\Delta}$. Hence a Fourier transform can be performed over the separation $R'-R$. This gives,
\begin{align}
H'_{\Delta'\Delta}(p) &\equiv \frac{1}{L}\sum_{R',R} e^{-ipR'} H_{R'R\Delta'\Delta} e^{ipR}\\
&= \frac{1}{L}\sum_{R',R} e^{-ip(R'-R)} H_{R'-R, 0, \Delta',\Delta}
\label{eq:FourierTransformEffHamiltonian}
\end{align}

Applying this to Eq.~\eqref{eq:EffExcHamRDeltaSpace} and adding the contribution from the $v$ hopping gives
\begin{equation}
\begin{aligned}
 H'_{\Delta'\Delta}(p) = \delta_{\Delta'\Delta}(2v + t_{00--} - t_{00++}) \\+t_{\Delta'\Delta--} -e^{ip(\Delta'-\Delta)} t_{\Delta'\Delta++}   
\end{aligned}
\end{equation}
This is the complete effective Exciton Hamiltonian for a two sublattice model with arbitrary hoppings. The effective Exciton Hamiltonian is calculated in an identical fashion for the Hubbard interactions $U, U'$ to give,
\begin{equation}
\begin{aligned}
H_{\Delta'\Delta}(p) = \delta_{\Delta'\Delta}\bigg[2v + U \delta_{\Delta \neq 0} + \frac{U'}{4} (\delta_{\Delta\neq1}+\delta_{\Delta \neq -1} -2) \\+t_{00--} - t_{00++} -\frac{U'}{2} \delta_{\Delta0} \cos{p}\bigg] \\+t_{\Delta'\Delta--} -e^{ip(\Delta'-\Delta)} t_{\Delta'\Delta++},
\label{eq:finalEffExcHam}
\end{aligned}
\end{equation}
where we use $\delta_{x\neq y} = 1- \delta_{x, y}$.
\subsection{Generic Hubbard Interaction Terms}

We now consider a generic translationally invariant Hubbard-like interaction term of the form,
\begin{equation}
\hat H_{int}^{r, i, j} = U^{r, i,j} \sum_{\tilde R} n_{\tilde R, i}n_{\tilde R+r, j} 
\end{equation}
for $i, j \in \{A, B\}$ where $U^{r, i, j}$ is the interaction strength between electrons at unit cell $\tilde R$, sublattice $i$ and those at unit cell $\tilde R+r$, sublattice $j$.
We first rewrite the interaction in the band basis. In this basis the number operator is rewritten as
\begin{equation}
n_{\tilde R, i} = \sum_{\alpha, \beta \in\{A, B\}} M_{i, \alpha} M_{i, \beta } c^\dagger_{\tilde R, \alpha} c_{\tilde R, \beta}
\end{equation}
Therefore the Hubbard term becomes,
\begin{multline}
\hat H_{int}^{r, i, j} = \sum_{\alpha, \beta, \gamma, \epsilon} M_{i, \alpha}M_{i, \beta}^*M_{j, \gamma}M^*_{j, \epsilon} c^\dagger_{\tilde R, \alpha} c_{\tilde R, \beta} c^\dagger_{\tilde R+r, \gamma} c_{\tilde R+r, \epsilon},
\end{multline}
for $\alpha, \beta, \gamma, \epsilon \in \{+, -\}$. 
To obtain the effective exciton Hamiltonian for this Hubbard term, the ground state expectation value is first calculated,
\begin{equation}
\begin{aligned}
\bra{GS} \hat H_{int}^{r, i, j} \ket{GS} &= U^{r, i,j} \sum_{\tilde R} \bra{GS} n_{\tilde R, i}n_{\tilde R+r, j}\ket{GS}\\
&= \frac{U^{r, i,j}}{4} \sum_{\tilde R} (1 + (2\delta_{i, j} -1)\delta_{\tilde R+r = \tilde R})\\
&= \frac{U^{r, i,j} L}{4} (1 + (2\delta_{i, j} -1) \delta_{r = 0}).
\end{aligned}
\end{equation}
The matrix elements of the effective exciton Hamiltonian are, 
\begin{equation}
\begin{aligned}
H_{R', R, \Delta', \Delta}^{r, i, j} =  \bra{GS} c^\dagger_{R', +}c_{R'+\Delta', -} \hat H_{int}^{r, i, j} c^\dagger_{R+\Delta, -} c_{R, +} \ket{GS} \\- \bra{GS} \hat H_{int}^{r, i, j}\ket{GS},
\end{aligned}
\end{equation} 
and can be calculated using Wick's theorem,
\begin{multline}
H_{R', R, \Delta', \Delta}^{r, i, j} = \frac{U^{r, i,j}}{4} \delta_{\Delta, \Delta'} \bigg\{\delta_{R, R'}\bigg[2(\delta_{r,0} - 1) \\+(\delta_{r\neq\Delta}+\delta_{r\neq-\Delta}) -2(2\delta_{i, j} -1)\delta_{r,0} \delta_{\Delta\neq 0}\bigg]\\ + (2\delta_{i, j} -1) \delta_{\Delta=0} (\delta_{R'+r, R}+\delta_{R', R+r})\bigg\}.
\end{multline}
Performing a Fourier transform as in Eq.~\eqref{eq:FourierTransformEffHamiltonian} gives,
\
\begin{multline}
H_{\Delta', \Delta}^{r, i, j}(p) = \frac{U^{r, i,j}}{4} \delta_{\Delta, \Delta'} \bigg[2(\delta_{r,0} - 1)\\+(\delta_{r\neq\Delta}+\delta_{r\neq-\Delta}) -2(2\delta_{i, j} -1)\delta_{r,0} \delta_{\Delta\neq 0}\\ + 2(2\delta_{i, j} -1) \delta_{\Delta=0}\delta_{\Delta'=0} \cos(rp)\bigg].
\label{eq:interactionEffHam}
\end{multline}
Note that the contributions to the effective exciton Hamiltonian from Hubbard interactions are only along the diagonal (\emph{i.e.} $\Delta= \Delta'$). This  is important for understanding why the pair-hopping term introduced in the main text (Eq.~\eqref{eq: full gapping term}) is necessary to gap out a nontrivial band. This is discussed in detail in App.~\ref{apdx:FullyGapped}. 

\section{Gapping out a nontrivial band}
\label{apdx:FullyGapped}

In this appendix we show that it is not possible to fully gap out a nontrivial exciton band (with nonzero exciton shift $s_\mathrm{exc} = 1/2$) using just hopping terms in the reduced $3\times 3$ Hamiltonian introduced in the main text, Eq.~\eqref{eq:3x3Matrix}. We begin with the final expression from App.~\ref{apdx:GeneralTheory} for the effective exciton Hamiltonian (equation \ref{eq:finalEffExcHam}). By translational symmetry, $t_{R,R',\alpha,\beta} = t_{R-R', 0,\alpha,\beta}$ and by $\mathcal{I}$ symmetry
$t_{R, R', \alpha\alpha} = t_{R', R, \alpha\alpha}$. The reduced $\Delta = -1, 0, 1$ space matrix therefore becomes, 
\begin{widetext}
\begin{equation}
H'(p)=
    \begin{pmatrix}
    2v +U - \frac{U'}{4} & (t_{-1, 0, --}-e^{-ip}t_{-1, 0, ++}) & (t_{-1, 1, --}-e^{-2ip}t_{-1, 1, ++}) \\
    (t_{-1, 0, --}-e^{ip}t_{-1, 0, ++}) & 2v-\frac{U'}{2} \cos(p) & (t_{0, 1, --}-e^{-ip}t_{0, 1, ++})\\
    (t_{-1, 1, --}-e^{2ip}t_{-1, 1, ++}) & (t_{0, 1, --}-e^{ip}t_{0, 1, ++}) & 2v +U - \frac{U'}{4}
    \end{pmatrix}.
    \label{eq:3x3MatrixGeneralHopping}
\end{equation}
\end{widetext}
We ignore the contributions from the hopping terms along the diagonal; these just give a constant shift in energy with no momentum ($p$) dependence and hence do not help in gapping out a nontrivial exciton band. At $p=0$ and $p=\pi$ this reduced effective Hamiltonian commutes with the $\mathcal{I}$ operator which, in this basis is,
\begin{equation}
U'_{\mathcal{I}} = \begin{pmatrix}
    0 & 0 & 1\\
    0 & 1 & 0\\
    1 & 0 & 0\\
\end{pmatrix}.
\label{eq:inversionDef3x3}
\end{equation}
Hence, the negative $\mathcal{I}$ state $(1, 0, -1)^T$ is an eigenstate at $p=0$ and $p=\pi$. The only terms in \ref{eq:3x3MatrixGeneralHopping} which couple to this eigenstate are $H_{\Delta=-1, \Delta'=-1}(p)$, $H_{\Delta=1, \Delta'=1}(p)$, $H_{\Delta=-1, \Delta'=1}(p)$ and $H_{\Delta=1, \Delta'=-1}(p)$. The diagonal terms [$H_{\Delta=-1, \Delta'=-1}(p)$ and $H_{\Delta=1, \Delta'=1}(p)$] have no momentum dependence and the off-diagonal terms [$H_{\Delta=-1, \Delta'=1}(p)$ and $H_{\Delta=1, \Delta'=-1}(p)$] are equal at $p=0$ and $p=\pi$. Hence the negative $\mathcal{I}$ eigenstate must have the same energy at $p=0$ and $p=\pi$. As a result, the nontrivial band, which must have negative $\mathcal{I}$ at only one of $p=0, \pi$, can never be fully gapped such that the top of the nontrivial band is lower in energy than the bottom of the remaining higher energy bands. The Hamiltonian above only includes the $U,U'$ nearest neighbour Hubbard-interaction terms (introduced in the main text). However, longer range Hubbard interaction terms also do not allow a nontrivial band to be fully gapped. From Eq.~\eqref{eq:interactionEffHam}  it can be seen that interaction terms can only give momentum dependent terms at $\Delta = 0, \Delta' = 0$. This term in the effective exciton Hamiltonian doesn't couple to the negative $\mathcal{I}$ eigenstate and hence it cannot provide split the energy of the negative $\mathcal{I}$ states at $p=0$ and $p=\pi$. Hence, longer range Hubbard interactions, much like longer range hoppings cannot be used to fully gap a nontrivial band.  This justifies our addition of a pair-hopping interaction term in Eq.~\eqref{eq: full gapping term} to fully gap out the shift exciton band.

\section{Flat Band Excitons}
\label{apdx:FlatBand}

\begin{figure}
    \centering
    \includegraphics[width=0.45\textwidth]{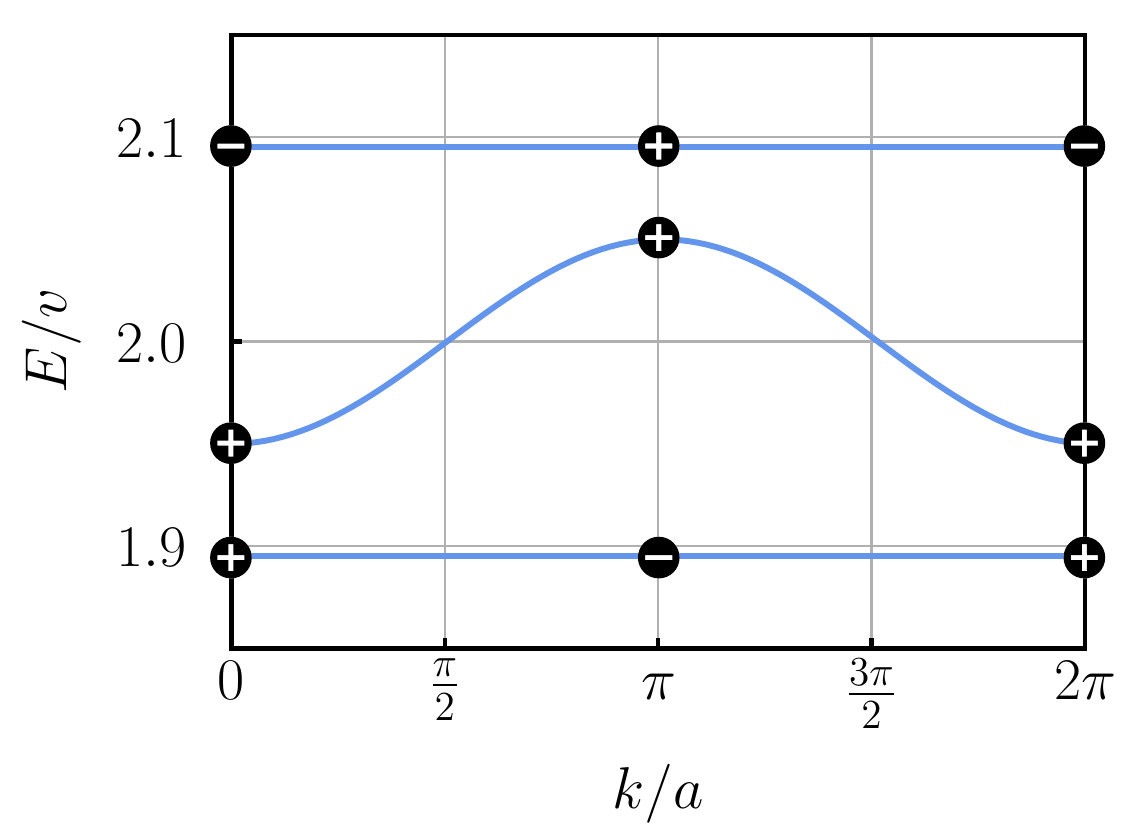}
    \caption{Flat nontrivial exciton band achieved using pair hopping terms. Parameters are, $v = 1.0$, $U = 0.02$, $U' = 0.1$, $w = 0.0$, $V = 0.1$ (for pair hopping term in \ref{eq:FlatBandPairHopping}). The inversion eigenvalues are labelled at $p=0,\pi$.}
    \label{fig: flatbandExcitons}
\end{figure}

Here we present the pair-hopping term required to obtain a fully gapped nontrivial exciton \emph{flat} band in the low energy $3\times3$ effective exciton Hamiltonian. We wish to find an effective Hamiltonian $H'(p)$ where $U'_{\mathcal{I}} H'(p)U_{\mathcal{I}}^{'\dagger} = H'(-p)$ (see Eq.~\eqref{eq:inversionDef3x3} for definition of $U'_{\mathcal{I}}$) and
\begin{equation}
H'(p) \psi(p) = \epsilon \psi(p), \qquad \forall p \in [0, 2\pi)
\end{equation}
where $U'_{\mathcal{I}} \psi(0) = \psi(0)$ and $U'_{\mathcal{I}} \psi(\pi) = -\psi(\pi)$. We choose an (arbitrary) explicit form for $\psi(p)$ which has these properties under $\mathcal{I}$ symmetry,
\begin{equation}
\psi(p) = \frac{1}{\sqrt{2}} \begin{pmatrix}
e^{ip} \\
0\\
1
\end{pmatrix}.
\label{eq:flatbandwavefunction}
\end{equation}
Requiring that this is an eigenvector with eigenvalue $\epsilon$ at all momenta, constrains the matrix elements of the $3\times3$ matrix $H(p)$. If the Hamiltonian $H(p)$ has elements $H_{\Delta,\Delta'}(p)$ then the eigenvector constraints are,
\begin{equation}
\begin{aligned}
&H_{1, -1}(p) e^{ip}+ H_{1, 1}(p) = \epsilon,\\
&H_{0,-1}(p) e^{ip} + H_{0, 1}(p) = 0, \\
&\bigl[\epsilon - H_{-1, -1}(p)\bigr]e^{ip} = H_{-1,1}(p).
\end{aligned}
\end{equation}

The matrix elements are further constrained by requiring that the effective exciton Hamiltonian commutes with $\mathcal{I}$ symmetry [\emph{i.e.} $U'_{\mathcal{I}} H'({p}) U_{\mathcal{I}}^{'\dagger} = H'({-p})$] hence,
\begin{equation}
\begin{aligned}
&H_{-1,-1}(p) = H_{1, 1}(-p),\\
&H_{-1,0}(p) = H_{1, 0}(-p),\\
&H_{-1,1}(p) = H_{1, -1}(-p).
\end{aligned}
\end{equation}
Finally, their are the constraints on the matrix elements arising from the Hermiticity of $H(p)$. This complete set of constraints can be used to determine the terms required to get a flat nontrivial flat band. One possible solution is to set all hopping terms to 0 and consider only the $U, U'$ interactions and a simple pair hopping term which takes the form,
\begin{multline}
H^{pair}_{\Delta, \Delta'}(p) = -V(\delta_{\Delta', -1}\delta_{\Delta, 1}  e^{-ip}\\+\delta_{\Delta', +1}\delta_{\Delta, -1} e^{ip})
\end{multline}
In the Hamiltonian before the projection in to the exciton basis this is a term,
\begin{multline}
\hat H^{pair} = -\frac{V}{4}\sum_{\tilde R} (c^\dagger_{\tilde R, B}c_{\tilde R+1, A}c^\dagger_{\tilde R, A}c_{\tilde R+1, B}\\+ c^\dagger_{\tilde R, B}c_{\tilde R-1, A}c^\dagger_{\tilde R, A}c_{\tilde R-1, B})
\label{eq:FlatBandPairHopping}
\end{multline}
The resulting spectrum, with nontrivial flat band at lowest energy, is plotted in Fig.~\ref{fig: flatbandExcitons}. Using Eq.~\eqref{eq:flatbandwavefunction}, the exciton Wannier state can be calculated explicitly for the flat band, 
\begin{equation}
W^R_{\Delta} = \frac{1}{\sqrt{2}}(\delta_{R=0} \delta_{\Delta = -1}+\delta_{R=1} \delta_{\Delta = 1}).
\end{equation}
This is compactly localised and obeys the symmetries we expected for the Wannier states of shift excitons [e.g. see Eq.~\eqref{eq:WanniershiftExcitons}].

\end{document}